\documentclass[preprint,10pt]{elsarticle}

\usepackage{multirow}

\usepackage{graphicx}
\usepackage{amssymb}

\journal{Vehicular Communications}

\begin{document}

\begin{frontmatter}


\title{A systematic mapping study on security countermeasures of in-vehicle communication systems
}



\author[l1,l2]{Jinghua Yu}
\ead{yujinghua@tongji.edu.cn}
\author[l2]{Stefan Wagner}
\ead{stefan.wagner@iste.uni-stuttgart.de}
\author[l1]{Bowen Wang}
\ead{bowen@tongji.edu.cn}
\author[l1]{Feng Luo}
\ead{luo_feng@tongji.edu.cn}

\address[l1]{School of Automotive Studies, Tongji University, Caoan Highway 4800, 201804 Shanghai, China}
\address[l2]{Institute of Software Engineering, University of Stuttgart, Universitätsstraße 38, 70569 Stuttgart, Germany} 

\begin{abstract}
The innovations of vehicle connectivity have been increasing dramatically to enhance the safety and user experience of driving, while the rising numbers of interfaces to the external world also bring security threats to vehicles. Many security countermeasures have been proposed and discussed to protect the systems and services against attacks. To provide an overview of the current states in this research field, we conducted a systematic mapping study on the topic area “security countermeasures of in-vehicle communication systems”. 279 papers are identified based on the defined study identification strategy and criteria. We discussed four research questions related to the security countermeasures, validation methods, publication patterns, and research trends and gaps based on the extracted and classified data. Finally, we evaluated the validity threats, the study identification results, and the whole mapping process.  We found that the studies in this topic area are increasing rapidly in recent years. However, there are still gaps in various subtopics like automotive Ethernet security, anomaly reaction, and so on. This study reviews the target field not only related to research findings but also research activities, which can help identify research gaps at a high level and inspire new ideas for future work.

\end{abstract}


\end{frontmatter}


\section{Introduction}
\label{S1}
The connected vehicle is a big trend in the automotive industry. A report from Juniper Research in 2018 predicted that 775 million consumer vehicles will be connected via telematics or by in-vehicle apps by 2023, rising from 330 million vehicles in 2018 \cite{Juniper2018} \cite{Upstream2020report}. Another report from Frost \& Sullivan indicates that connected vehicles will comprise nearly 86\% of the global automotive market by 2025 \cite{Frost2020Sullivan} \cite{Upstream2021report}. With various cyber interfaces, users can receive external information like entertainment streams, real-time traffic conditions, or the nearest gas station’s location to increase the user experience and driving safety. The vehicle can install or upgrade software or firmware remotely to add new features or fix bugs. Furthermore, the data can also be sent from the vehicle side. An SOS message can be triggered immediately for rescue when an accident happens. User’s customized data can be uploaded to the cloud for backup or management. The vehicle’s operation states can be collected by the manufacturers for product improvement or after-sale services. More service models will be continuously innovated with this exciting property.

However, intelligent connectivity also brings cybersecurity threats to stakeholders. The Upstream research team analyzed 633 publicly reported incidents since 2010, while there were 207 incidents in 2020 (by Nov. 25, 2020) \cite{Upstream2021report} and 155 in 2019 (by Dec. 7, 2019) \cite{Upstream2020report}. The majority of the attackers are black hat (54.6\% in 2020) \cite{Upstream2021report}, who hack the system for personal gain or malicious purposes and cause losses to the stakeholders. Therefore, it's increasingly important and urgent to keep an eye on the security issues of modern cars. 

The in-vehicle communication system, as our main research interest, has many attack surfaces like ECU (Engine Control Unit) / TCU (Telematics Control Unit) / GW (Gateway) and in-vehicle networks, which are raked as the 8th (4.27\%) and 9th (3.75\%) common attack vectors in the Upstream 2021 report \cite{Upstream2021report}. Researchers have proposed various countermeasures to protect the target systems against hacking attempts and also conducted surveys to investigate current research states of existing security countermeasures. However, no systematic mapping study has been published according to our knowledge. They mainly investigated the concrete solutions but omitted the research activity patterns at a more general level (e.g. the research year and publication venue), which can reveal more information of the research state and inspire new research ideas.

To better understand the current research states and provide an overview of the target research field, we conducted a systematic mapping study on the topic: “security countermeasures of in-vehicle communication systems”. We raised four research questions related to the research content and activities, designed concrete working protocols, and conduct the study accordingly. We analyzed and classified 279 included articles, answered the research questions with statistical evidence, and provided the mapping with categories for this topic area. Finally, we evaluated this mapping study from the perspective of the validity threats, the study identification, and the whole mapping process.

The rest of this article is organized as follows. In the “Related Work” section, we summarize the existing reviews on this topic and introduce three review methods and their differences. In the “Study Process” section, we introduce the overall workflow including work strategies and criteria and present the real work process of study identification and data extraction and classification. In the “Data Visualization and Analysis” section, we classify the data and visualize them with (stacked) bar or bubble charts. Four research questions are answered based on the visualized data. In the “Study evaluation” section, we discuss the validity threats in five aspects and evaluate the study identification and the whole mapping process. Finally, we summarize this article in the “Conclusion” section.

\section{Related Work}
\label{S2}

\subsection{Existing reviews}
\label{S2.1}

Several review articles have been published in this topic area. Hu and Luo \cite{hu2018review} discussed the state-of-the-art techniques of secure communication for in-vehicle networks including message authentication, data encryption, and intrusion protection, and concluded technical requirements of cryptographic mechanisms and intrusion detection policy. Zoppelt and Kolagari \cite{zoppelt2019today} described seven typical automotive security and dependability mechanisms, examined them with a set of known attacks, and led to the conclusion that most countermeasures against attacks are hardly effective. 

Other than the reviews at a general level, some researchers focus on specific techniques. Loukas et. al. \cite{loukas2019taxonomy} conducted a survey on the intrusion detection systems (IDS) of vehicles, proposed a classification scheme, and discussed learned lessons and open issues in relevant industries. Lokman, Othman, and Abu-Bakar \cite{lokman2019intrusion} investigated the IDS implementation of the CAN network systems in four aspects (i.e. detection approaches, deployment strategies, attacking techniques, and technical challenges) and categorized them based on seven detection strategies. Grimm, Pistorius, and Sax \cite{grimm2020network} presented a classification of network monitoring techniques for security purposes and compared typical security measures in enterprise information technology systematically with the ones in the vehicle security field.

Additionally, some reviews are on a certain type of communication media. Gmiden et. al. \cite{gmiden2019cryptographic} and Bozdal et. al. \cite{bozdal2020evaluation} discussed the vulnerabilities of the CAN bus protocol, presented and compared various countermeasures against attacks. Kishikawa et. al.\cite{kishikawa2019vulnerability} highlighted the vulnerabilities of the FlexRay and discussed countermeasures against spoofing attack on FlexRay networks. Boatright and Tardo \cite{boatright2012security} examined the security aspects of the Ethernet AVB backbone and discussed various Ethernet security solutions including protocols, algorithms, and encryption mechanisms appropriate for vehicular requirements.

All mentioned reviews are not systematic literature reviews or systematic mapping studies and mainly focus on the research findings with no attention to research activities.

\subsection{Review methods}
\label{S2.2}

Reviewing is an essential research method, which searches, collects, summarizes, and synthesizes existing studies to present current knowledge and research states of a research area. Literature review (LR), systematic literature review (SLR), and systematic mapping study (SMS) (or called scoping study, mapping review) are three common methods for review studies. 

LR summarizes evidence on a topic using informal or subjective methods to collect and interpret studies qualitatively to provide a summary or overview of a topic \cite{kysh2013s}. This kind of review can be found in the introduction or related work sections of research articles or in some review articles. 

Due to the informality and subjectivity of LR, SLR is a better choice to conduct a review with low biases. SLR is a high-level overview of primary research that identifies, selects, synthesizes, and appraises all relevant high-quality research evidence and eliminates bias \cite{kysh2013s}.

Although SLR is good at answering focused questions and providing evidence with low biases, it does not present an overview of a topic area. By contrast, an SMS is a defined method to build classification schemes and structure a research field \cite{petersen2008systematic} to discover research trends and gaps \cite{petersen2015guidelines}. SMS is not only based on the findings presented in publications but also the activities related to the findings \cite{cooper2016mapping} like when and where a paper was published. The outcome of an SMS is an inventory of papers on the topic area which is mapped to a classification scheme \cite{petersen2015guidelines}. 

SMS was originally used in medical fields and then started to get more attention in software engineering around 2007 \cite{petersen2008systematic}. Petersen et al. \cite{petersen2008systematic} provided a guideline to conduct an SMS in software engineering in 2008, which is the most used SMS guidelines in software engineering \cite{petersen2015guidelines}. In 2015, Petersen et al. \cite{petersen2015guidelines} conducted an SMS of SMS to identify improvement potentials in conducting the mapping studies and proposed an updated guideline for conduction SMS in software engineering. 

Although the mentioned SMS guidelines and studies are proposed in software engineering fields, the SMS processes are independent of software engineering and are also applicable to other research areas. For example, Zanella Gomes et al. \cite{zanella2019ubiquitous} conducted a survey on the provisioning of ubiquitous intelligent services for vehicular users and identified clusters of research interest on this topic. Lun et al. \cite{lun2019state} focused on how security is addressed in the cyber physical systems from an automatic control perspective and provided a systematic map of this subject. Both examples are conducted based on the systematic process proposed in \cite{petersen2015guidelines}. In this article, we report a systematic map on the security countermeasures of the in-vehicle communication systems to present current research states and identify trends and gaps in this topic area by using the updated guideline proposed by Petersen et al. \cite{petersen2015guidelines}.

\section{Study Process}
\begin{figure}[ht]
\centering\includegraphics[width=1.0\linewidth]{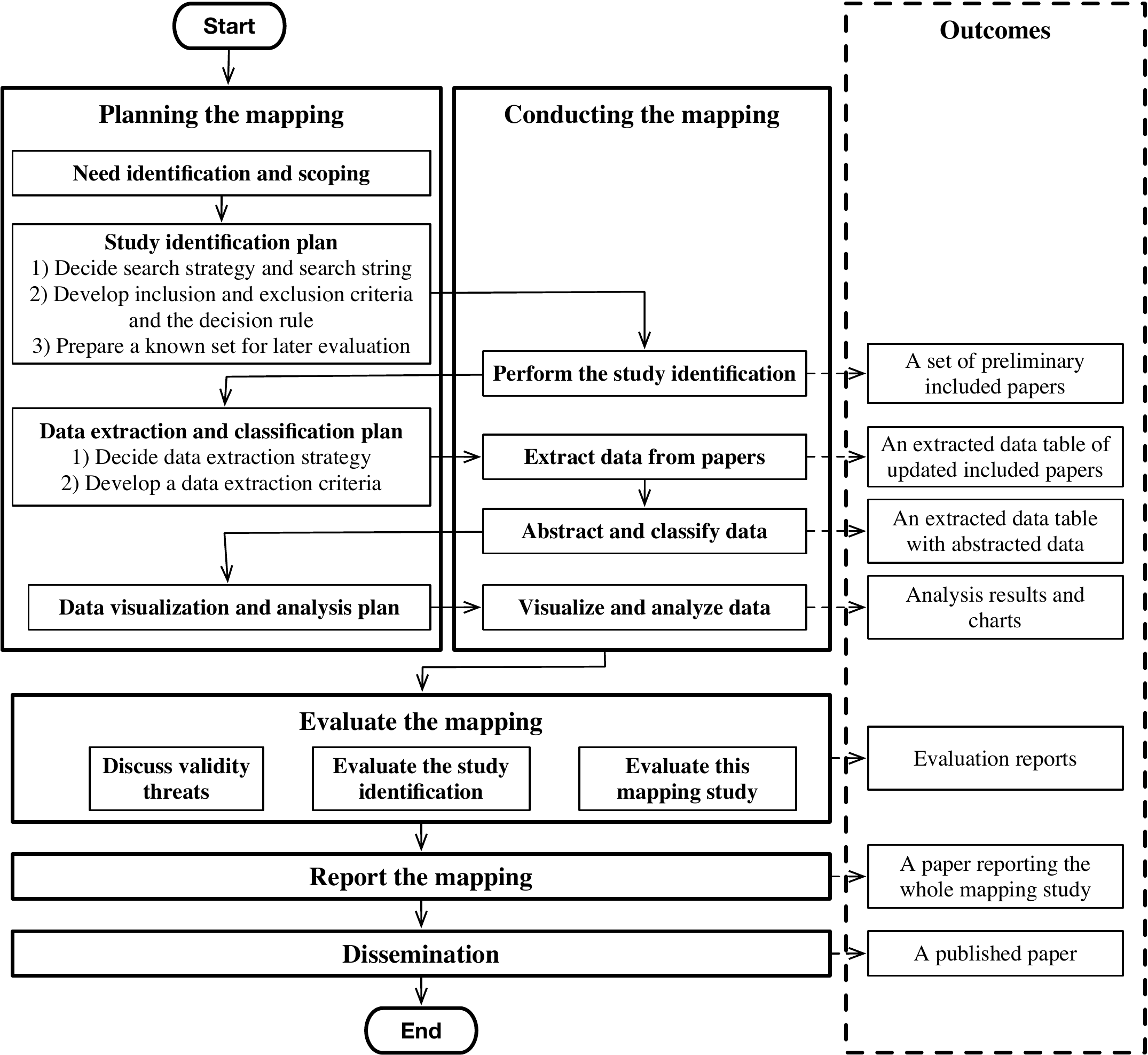}
\caption{Overall Workflow of the Study}
\label{fig_overallWorkflow}
\end{figure}

\label{S3{}}
The overall workflow and activities applied in this study are shown in Figure \ref{fig_overallWorkflow}. Five main activities are involved in the study process, which are planning, conducting, evaluating, reporting, and dissemination. In this section, we mainly follow the workflow to elaborate the study activities with corresponding outcomes except the results of data analysis and mapping evaluation, which will be discussed in the section \ref{S4} and \ref{S5} respectively.

\subsection{Need identification and scoping}
\label{S3.1}
The goal of this study is to summarize the current research states and identify the trends and gaps in the target research area, which is the security countermeasures of in-vehicle communication systems. This study also serves as a preliminary investigation for further systematic literature reviews on a particular security issue or solution.

Four research questions (RQ) are proposed as follows.

\begin{itemize}
\item RQ1: What countermeasures were proposed or evaluated for in-vehicle communication systems?
\item RQ2: How were the proposed or existing countermeasures evaluated?
\item RQ3: When and where were related studies conducted and published? 
\item RQ4: What are the research trends and gaps in this topic area?
\end{itemize}

\subsection{Study identification plan}
\label{S3.2}

\subsubsection{Search strategy}
\label{S3.2.1}

We chose five commonly used databases in the automotive domain for the search, which are IEEE Xplore, ACM Digital Library, Web of Science, Engineering Village, and SAE Mobilus. The search string was first identified from the target topic and research questions, and then improved iteratively and dynamically by conducting several pilot searches. The final search string we used is “(secur* AND (countermeasure OR solution OR mechanism OR strateg* OR evaluation)) AND ((in-vehicle OR automotive OR automobile) AND (network OR communication))”. A set of already known papers was also prepared from the previous related research of authors for evaluating the study identification process in the later phase.

\subsubsection{Inclusion and exclusion criteria}
\label{S3.2.2}

To identify relevant papers from the preliminarily found publications, the inclusion and exclusion criteria should be defined. We specified the criteria based on the recommendations in the guideline, which are listed in Table \ref{table_in_ex_criteria}. During the inclusion and exclusion process, we found that the boundary of some target systems is ambiguous. Therefore, we identified an additional principle to solve the problem, which is: if an ambiguous system (e.g. anti-theft systems, over-the-air software update, wireless sensor networks, etc.) is related to in-vehicle network parts, it should be included; otherwise, it should be excluded. 

\begin{table}[ht]
\renewcommand{\arraystretch}{1.1}
\caption{Inclusion and Excluaion Criteria}
\label{table_in_ex_criteria}
\centering
\begin{tabular}{p{1.8cm} p{4.8cm} p{4.8cm}}
\hline
\textbf{Aspect} & \textbf{Inclusion Criteria} & \textbf{Exclusion Criteria}\\
\hline
Language and year & 1) English; 2) From the year 2010 to 2020
& 1) Not written in English; 2) Out of the defined period \\
\hline
Topic scope & Topics mainly related to the topic area, including 1) countermeasure, strategy, mechanism, solution; 2) design approaches, processes, architectures, frameworks; 3) solution evaluation (not security system evaluation)
for the in-vehicle communication systems to defense the attacks & Topics not mainly related to the topics area, including 1) threat analysis and risk management; 2) attack analysis; 3) discussion on issues or challenge; 4) general research not specifically for in-vehicle systems; 5) only mention the topic area as background; 6) security test, verification, evaluation\\
\hline
Target object scope & Main objects in in-vehicle communication systems, including 1) links, networks, communications; 2) Software or hardware related to in-vehicle communication systems & Objects out of the physical boundary of a vehicle, including 1) V2X, ad-hoc networks (VANET); 2) robotic vehicles \\
\hline
Paper type and availability & 1) Peer-reviewed articles; 2) Full text is available & 1) Short papers; 2) Grey literature; 3) Full text is not available \\
\hline
\end{tabular}
\end{table}

\subsection{Performing the study identification}
\label{S3.3}
\begin{figure}[ht]
\centering\includegraphics[width=0.8\linewidth]{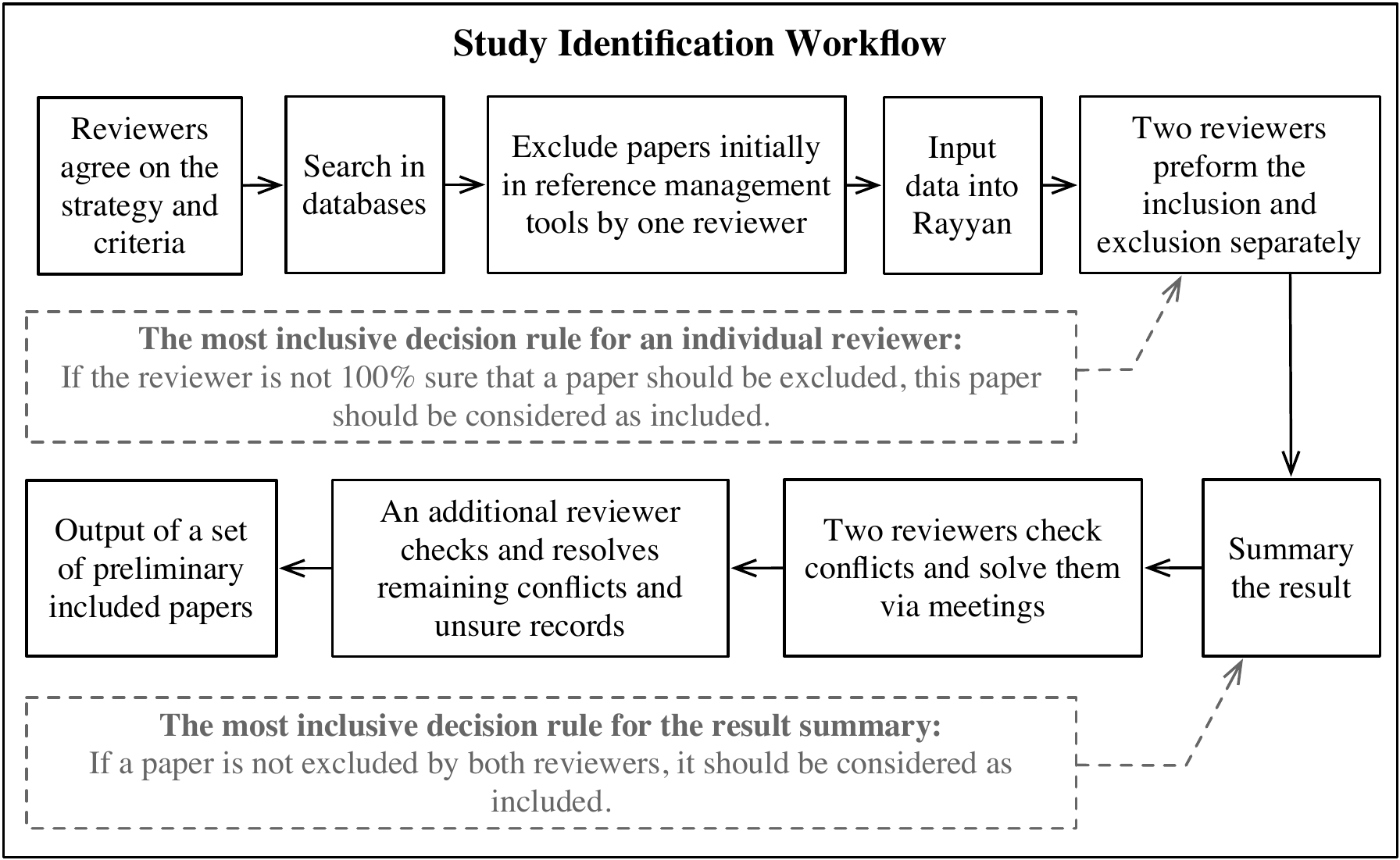}
\caption{Workflow of the Study Identification}
\label{fig_studyIdfWorkflow}
\end{figure}

The detailed workflow of the study identification is shown in Figure \ref{fig_studyIdfWorkflow}.

All reviewers should first discuss and agree on the study identification strategy and criteria. After the search, publications that are obviously duplicated or violating the inclusion criteria were removed using reference management tools (i.e. Zotero and Mendeley). Then, papers’ meta-data was input into a screening tool called Rayyan \cite{ouzzani2016rayyan} for further decision. Two reviewers did the inclusion and exclusion work separately according to the defined criteria. The most inclusive decision rule was defined to specify the behavior of an individual reviewer, which is: if the reviewer is not completely sure that a paper should be excluded, this paper should be considered as included. Similarly, the most inclusive rule for the result summary was defined, which is: if a paper is not excluded by both reviewers, it should be considered as included. When a decision conflict occurred, the two reviewers discussed with each other to solve it. If it failed to reach an agreement, an additional reviewer was involved and made the final decision. 

Note that the decision in this step is based on the metadata of papers (i.e. title, keywords, and abstract). The output list of the included paper is preliminary. The final decision would be made based on the full text in the paper extraction process.

\begin{figure}[ht]
\centering\includegraphics[width=0.8\linewidth]{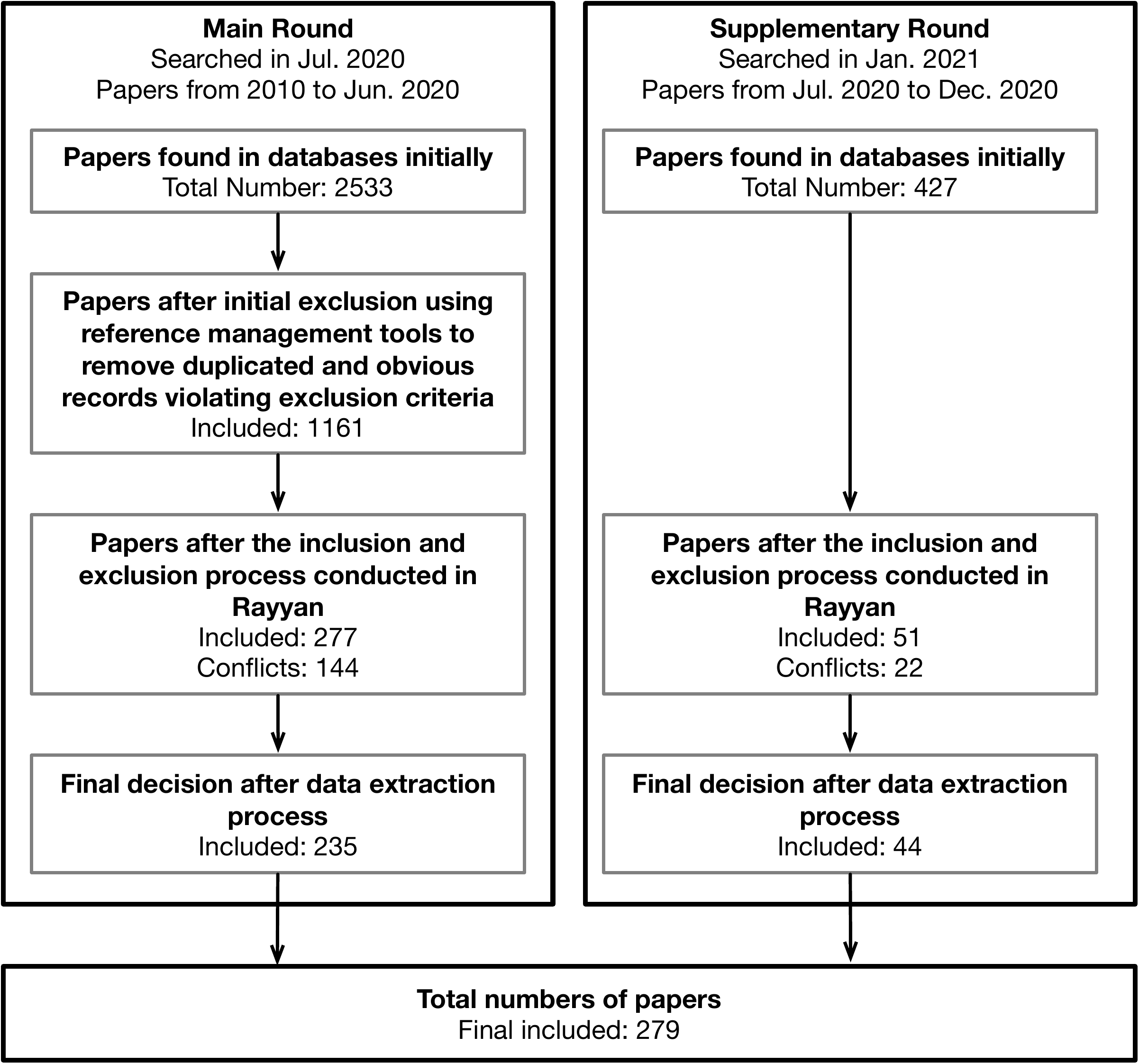}
\caption{Paper number records in the study identification process}
\label{fig_paperNumRecords}
\end{figure}

Figure \ref{fig_paperNumRecords} shows the paper numbers in the study identification process. In practice, we did the search in two rounds due to the time schedule. The main round started in July 2020 to identify and analyze papers from 2010 to Jun. 2020, and the supplementary search was performed in Jan. 2021 to include the latest papers published in the second half of 2020.

\subsection{Data extraction and classification plan}
\label{S3.4}

\subsubsection{Data extraction strategy}
\label{S3.4.1}
\begin{figure}[ht]
\centering\includegraphics[width=0.8\linewidth]{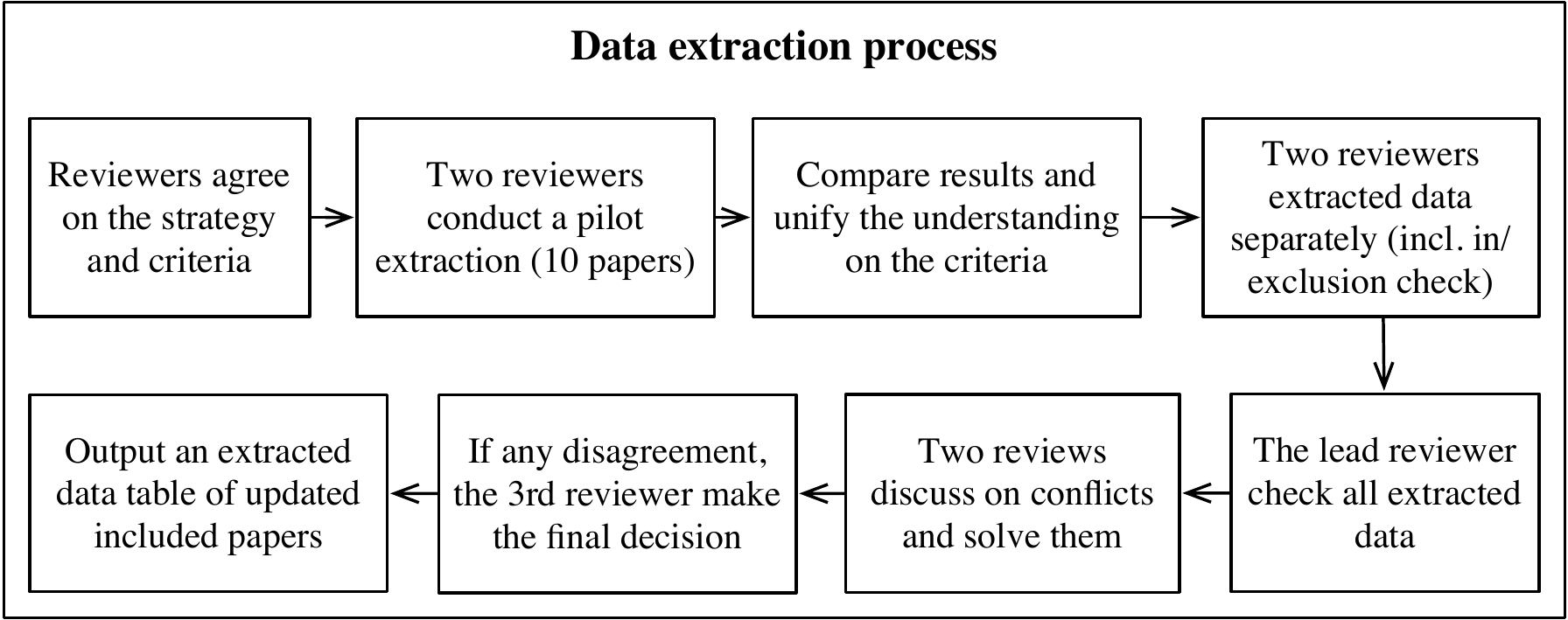}
\caption{Data Extraction Process}
\label{fig_dataExtraProcess}
\end{figure}

To extract data effectively and efficiently, We first discussed and developed a data extraction strategy to specify the process (shown in Figure \ref{fig_dataExtraProcess}).

After agreeing on the extraction strategy and criteria, two reviewers first extracted data of a small set of papers (10 papers) separately and compared the results to make the understanding of the extraction criteria consistent. Then, reviewers started the extraction of different set papers to ensure work efficiency. Papers were also checked according to the inclusion and exclusion criteria in the extraction process. We mainly got data from the paper abstract, conclusion, contribution, section titles, tables, and figures to obtain the required information efficiently. The extracted results were checked by the lead reviewer again to ensure the extraction quality. If there is any unsure question, the reviewers held a meeting to solve it. The 3rd reviewer was involved to make the final decision for conflicts. Finally, a table with extracted data of included papers was output.

\subsubsection{Data extraction criteria}
\label{S3.4.2}

The data extraction criteria were developed in the form of an extraction template. The topic-independent and topic-dependent extracted items were specified with descriptions and related research questions in Table \ref{table_TopicIndExtraTemplate} and \ref{table_TopicDepExtraTemplate} respectively. The values of extracted items can be divided into enumerated and unenumerated ones. The enumerated values (EV), like countries and publication years, are limited or defined in advance, which can be counted directly in later phases. The unenumerated values (UEV) are open and can not be classified before the extraction, like countermeasures and properties concerned. Further abstraction and classification are required after the extraction.

\begin{table}[ht]
\renewcommand{\arraystretch}{1.1}
\caption{Topic-independent Extracted Data Template}
\label{table_TopicIndExtraTemplate}
\centering
\begin{tabular}{p{1.8cm} p{7.2cm} p{1.2cm}}
\hline
\textbf{Extracted Items} & \textbf{Descriptions} & \textbf{Related RQs}\\
\hline
Paper ID & [Basic Information] Assigned ID for the paper & / \\
\hline
Paper title & [Basic Information] Name of the paper & / \\
\hline
Author & [Basic Information] Author(s) of the paper & / \\
\hline
Length & [Basic Information] Page number of the paper & / \\
\hline
Research Type & [Basic Information] The major research type of the paper according to the classification scheme proposed by Wieringa et al. \cite{wieringa2006requirements} & / \\
\hline
Country & [EV] Authors’ countries according to affiliations (record country code) & RQ3 \\
\hline
Author’s Field & [EV] Whether the authors are from academia(A), industry(I) or both(B). The research institute belongs to the academia.  & RQ3 \\
\hline
Year & [EV] The year of the publication (from 2010 to 2020) & RQ3 \\
\hline
Publication Type & [EV] Whether the paper is a journal article (JA) or a conference article (CA) & RQ3 \\
\hline
Venue & [UEV] In which journal or conference was the paper published. & RQ3 \\
\hline

\end{tabular}
\end{table}

\begin{table}[ht]
\renewcommand{\arraystretch}{1.1}
\caption{Topic-dependent Extracted Data Template}
\label{table_TopicDepExtraTemplate}
\centering
\begin{tabular}{p{1.8cm} p{7.2cm} p{1.2cm}}
\hline
\textbf{Extracted Items} & \textbf{Descriptions} & \textbf{Related RQs}\\
\hline
Main Countermeasure & [UEV] Main referred countermeasure(s) of the paper & RQ1 \\
\hline
Media Type & [EV] The type of the communication media in the target system, including CAN, CAN FD, FlexRay, IP-based, others, and N/A (no specific media type strictly related to the solution) & RQ1 \\
\hline
Property Concerned & [UEV] System properties concerned in the validation of solutions & RQ2 \\
\hline
Validation or Evaluation Method & [EV] Approach used to validate or evaluate the solution according to the classification scheme presented in the guideline \cite{petersen2015guidelines} & RQ2 \\
\hline
Validation Tool & [EV] Whether special hardware (HW) tools (e.g. an FPGA board) are used to validate solutions except computers. The enumerated values recorded here are PC, HW or none. & RQ2 \\
\hline
Security Goal & [UEV] Security goals, like confidentiality and availability, or attacks against by the solutions are recorded & RQ1 \\
\hline
Future Work & [UEV] The future work mentioned in the paper & RQ4 \\
\hline
\end{tabular}
\end{table}

For the research type, We applied the classification scheme proposed by Wieringa et al. \cite{wieringa2006requirements}. Six types of research with explanations are listed below. 

\begin{itemize}
\item Evaluation Research (ER): It is the investigation of a problem or an implementation of a technique in practice \cite{wieringa2006requirements}. We regarded surveys and reviews as this type.
\item Proposal of Solution (PS): It proposes a solution technique and argues for its relevance, without a full-blown validation, which must be novel, or at least a significant improvement of an existing technique \cite{wieringa2006requirements}. It should mainly focus on the solution itself.
\item Validation Research (VR): It investigates the properties of a solution proposal that has not yet been implemented in practice. The solution may have been proposed somewhere else \cite{wieringa2006requirements}. 
\item Philosophical Paper (PP): It sketches a new way of looking at things, a new conceptual framework, etc. \cite{wieringa2006requirements}. 
\item Opinion Paper (OP): It contains the author's opinion about what is wrong or good about something, how we should do something, etc. \cite{wieringa2006requirements}. 
\item Personal Experience Paper (PEP): It describes the personal experience and emphasize what and not why \cite{wieringa2006requirements}. 
\end{itemize}

There are only evaluation research (ER), proposal of solution (PS) and validation research (VR) included in this study. A significant difference between the validation and evaluation is that the validation is done in the lab, while the evaluation studies take place in real-world context \cite{petersen2015guidelines}. Note that if a paper proposed a novel solution and also presented detailed evaluation or validation research, we considered it as a PS paper.

For the validation or evaluation methods, a classification scheme has also been presented in Petersen et al.’s guideline \cite{petersen2015guidelines}. The methods of the validation research are stimulation as an empirical method, laboratory experiments (machine or human), prototyping, mathematical analysis and proof of properties, and academic case studies (e.g with students). The methods of the evaluation research include industrial case study, controlled experiments with practitioners, practitioner targeted survey, action research, and ethnography. 

\subsection{Performing extraction and classification}
\label{S3.5}
According to the data extraction strategy and criteria defined in section \ref{S3.4}, we extracted data of each included paper and recorded values in a table. Then, we abstracted the unenumerated values and developed classification schemes for them. 

For the values of the main countermeasure, nine abstracted countermeasures are identified, which are listed as follows. 

\begin{itemize}
\item Anomaly detection: Anomaly or intrusion detection methods of a system, e.g. a specification-based or machine-learning-based algorithm to detect the anomaly.
\item Anomaly reaction: Strategies or methods of system reaction when anomaly or intrusion has been detected, e.g. a method to disable the attacker when being detected.
\item Hardware support for security: Research on hardware to support better security countermeasures, e.g. a hardware accelerator for cryptographic calculation, realize a secure protocol in hardware.
\item Key management: Approach for key management in security mechanisms, e.g. key exchange protocols, methods of generating and storing keys, ect.
\item Secure communication scheme: Protocols or mechanisms to ensure the security of the data transmission in various networks, e.g. secure communication protocol, ID shuffling mechanisms, etc.
\item Secure component design: Security design methods for a single component in a system, e.g. a secure gateway, secure ECU design, etc.
\item Security mechanism at application layers: Mechanisms that are specific for a type of applications, e.g. a secure protocol for Unified Diagnostic Service in vehicles, a secure strategy for braking dynamics of cooperative driving, etc.
\item System security design: System-level design approaches like overall design architecture, framework, etc., e.g. distributed firewall deployment, a comprehensive framework including a set of countermeasures to protect the system.
\item Others: Other topics that are relevant but can not be classified into the rest classes, e.g. a study of bit level permutation which is used in the cryptographic environment.
\end{itemize}

For the properties concerned, we recorded many metrics like error rate and average delay originally from the papers. Then, we divided them into five classes, which are introduced below.

\begin{itemize}
\item effectiveness: Whether the solution is effective to solve the problem. Concrete metrics in this class include detection accuracy metrics (e.g. precision, recall, true positive, true negative, false positive, false negative, f-score), output randomness, etc.
\item efficiency: Whether the solution is efficient. Typical metrics in this class are latency, detection speed, round trip time, standby time, process throughput, etc.
\item performance: The performance metrics of the system include network throughput, busload, bandwidth, and metrics related to the system quality (e.g QoS, robustness, reliability, etc.). This class investigates the impacts of the countermeasures on the original system.
\item resource: This class represents the resource required by the proposed solution, like financial or computational cost, memory consumption, etc.
\item compatibility: Whether the solution is compatible with other contexts, e.g. be applicable for other media types, be compatible for other protocols.
\end{itemize}

For the validation or evaluation approach, we adapted the original classification to better fit our study. All research in PS and VR papers used validation methods to validate solutions, which means that they were all conducted in laboratory environments (including stimulations on PCs and experiments in the real-world). Two kinds of abstracted methods for the validation are “prototyping and then doing experiments (PE)” and “Theoretical analysis/proofing (TA)”.

For the security goals, we applied nine security properties classes proposed in the EVITA project \cite{ruddle2009deliverable}, which are data origin authenticity, integrity, authorization, freshness, non-repudiation, privacy, anonymity, confidentiality, and availability. Since many papers do not mention the security goals explicitly, we developed a mapping (Table \ref{table_sec_goal_attack_map}) to map the possible attacks or common mechanisms mentioned in publications to the nine security goals. If there is no specific security goals or related attacks explicitly mentioned in the paper, the security goal of this paper is marked as “None”.

\begin{table}[ht]
\renewcommand{\arraystretch}{1.1}
\caption{Map of Security Goals and Related Attacks or Mechanisms}
\label{table_sec_goal_attack_map}
\centering
\begin{tabular}{p{2.4cm} p{8cm} }
\hline
\textbf{Security Goal} & \textbf{Relevant attack or mechanism} \\
\hline
(Data Origin) Authenticity & Forging or replacing frames, spoofing, fuzzing attack, masquerading frames, message injection, cloaking attack, fabrication, man-in-the-middle attack \\
\hline
Integrity & Manipulation, tampering, modifying data fields \\
\hline
Authorization & Access control, unauthorized access and usage \\
\hline
Freshness & Replay attack \\
\hline
Non-repudiation & N/A. No study in this class was found. \\
\hline
Privacy & N/A. Studies explicitly mentioned privacy were classified in this class. \\
\hline
Anonymity & N/A. No study in this class was found. \\
\hline
Confidentiality & Sniffing, reverse engineering, inspectation, message encryption, eavesdropping \\
\hline
Availability & Denial-of-Service (Dos) attack, interruption, suspension attack, message dropping \\
\hline
\end{tabular}
\end{table}

For future work, we abstracted the next-step work mentioned by authors and divided them into categories, which are listed below.

\begin{itemize}
\item Improving performance: To improve the performance of the proposed solutions, including improving the practicability or effectiveness, implementing HW modules, overcoming current limitations, reducing cost, etc.
\item Extend application scope: To extend application scope for more scenarios, including fighting against or detecting more attacks, integrating the proposed solution with other protocols, supporting in other contexts, trying other HWs, etc.
\item Further evaluation: To evaluate the proposed solution under other conditions, including with a more extensive set of attacks, with other parameters, with real-world data and scenarios, with large scale, etc., or to investigate and compare the solution with other relevant ones.
\item Add features: To add new functional features to the proposed solution or to integrate the proposed solution with other methods, like adding a reaction process after the anomaly detection, etc.
\item Exploit other methods: To find other possible solutions for the mentioned issues or other problems, which is completely different from the proposed solution.
\item None: No concrete future work is mentioned in the paper.
\end{itemize}

After we defined all the abstraction and classification, we updated the extraction table based on the classification schemes and output a classified data table for further analysis.

\subsection{Data visualization and analysis plan}
\label{S3.6}
To visualize and analyze the extracted data, we chose the bubble chart and the (stacked) bar chart for data visualization.  These kinds of charts can display the relations between two or three variants and are the most commonly used means for the visulization \cite{petersen2008systematic}. We developed R scripts for efficient data statistics and drawing the plots. The detail of the data visualization and analysis is presented in section \ref{S4}.

\subsection{Evaluating the mapping}
\label{S3.7}

We evaluate the study from three aspects. First, the validity threats are discussed, including the potential biases and countermeasures we conducted to migrate the biases. Then, the identified study set is compared with a known paper set to evaluate the study identification. Finally, the whole mapping study process is evaluated based on the activities involved according to the evaluation rubric proposed by the guideline \cite{petersen2015guidelines}. The detail of the evaluation is presented in section \ref{S5}

\subsection{Reporting and disseminating the mapping}
\label{S3.8}

Writing a report and publishing it is the final step of an SMS. Mapping strategies, conducting details, and relevant discussion should be reported clearly. We used the workflow (Figure \ref{fig_overallWorkflow}) as a checklist to ensure that we do not miss any important information in the report. As for the venue, we would prefer a journal for the publication due to the space limits of conference papers. The study in \cite{petersen2015guidelines} also found a higher number of mapping studies published in reputable journals rather than conferences.

\section{Data Visualization and Analysis}
\label{S4}
\subsection{Data statistic and visualization}
\label{S4.1}

To conclude meaningful information from the extracted data, we used five keywords to guide the data statistic and visualization process, which are “year”, “country”, “countermeasure”, “validation approach”, and “PS”. The “year” charts display the paper distribution related to the publication time and present the publication profile throughout the past 10 years. The “country” charts show the distributions related to authors’ countries and fields. The “countermeasure” charts display the distribution related to the abstracted countermeasures of papers and represent the research interests in this topic area. The “Validation approach” charts show the validation approach distributions used in the PS and VR papers, including abstracted approaches, tools, and properties concerned in the validation. The “PS” charts show the main security goals and future work of the PS type papers.

Table \ref{table_chart_plan} summarizes the detailed information about the planned charts. We set a rule for the data statistic in this step. If a paper includes two target elements, it should be considered as two records and counted twice. For example, if a paper has two authors from two different countries, this paper was counted twice and belongs to two countries separately when counting countries in the statistic.

\begin{table}[ht]
\renewcommand{\arraystretch}{1.1}
\caption{Summary of Planned Charts}
\label{table_chart_plan}
\centering
\begin{tabular}{p{1.3cm} p{7.7cm} p{0.9cm} p{1.2cm}}
\hline
\textbf{Key-word} & \textbf{(Chart No.) Variant 1 - Variant 2 [ - Variant 3]} & \textbf{Chart Type} & \textbf{Figure No.} \\
\hline
\multirow{4}{1.5cm}{Year} & (1.1) numbers of papers - publication types - years & stacked bar & Fig. \ref{plot_1_1_pubType_year} \\
\cline{2-4} & (1.2) top journals - numbers of papers & bar & Fig. \ref{plot_1_2_journal} \\
\cline{2-4} & (1.3) top conferences - numbers of papers & bar & Fig. \ref{plot_1_3_conference} \\
\cline{2-4} & (1.4) media type - years - numbers of papers & bubble & Fig. \ref{plot_1_4_media_year} \\
\hline
\multirow{3}{1.5cm}{Country} & (2.1) top 10 countries - years - numbers of papers & bubble & Fig. \ref{plot_2_1_country_year} \\
\cline{2-4} & (2.2) numbers of papers - authors’ fields - top 10 countries & stacked bar & Fig. \ref{plot_2_2_country_field} \\
\cline{2-4} & (2.3) media type - top 10 countries - numbers of papers & bubble & Fig. \ref{plot_2_3_country_media} \\
\hline
\multirow{2}{1cm}{Counter-measure} & (3.1) countermeasures - research types - paper numbers & stacked bar & Fig. \ref{plot_3_1_coun_restype} \\
\cline{2-4} & (3.2) countermeasures - media types - papers numbers & bubble & Fig. \ref{plot_3_2_coun_media} \\
\hline
\multirow{4}{1.5cm}{Validation approach} & (4.1) PS validation approaches - validation tools - numbers of papers & stacked bar & Fig. \ref{plot_4_1_PS_valAppr_valTool} \\
\cline{2-4} & (4.2) VR validation approaches - validation tools - numbers of papers & stacked bar & Fig. \ref{plot_4_2_VR_valAppr_valTool} \\
\cline{2-4} & (4.3) PS properties concerned - numbers of papers & bar & Fig. \ref{plot_4_3_PS_valPropert} \\
\cline{2-4} & (4.4) VR properties concerned - numbers of papers & bar & Fig. \ref{plot_4_4_VR_valPropert} \\
\hline
\multirow{2}{1.5cm}{PS} & (5.1) countermeasures - main security goals - numbers of papers & bubble & Fig. \ref{plot_5_1_coun_goal} \\
\cline{2-4} & (5.2) future work - numbers of paper & bar & Fig. \ref{plot_5_2_PS_futurework} \\
\hline
\end{tabular}
\end{table}

\subsection{Answering research questions}
\label{S4.2}

With the help of the charts generated based on the data visualization plan in Table \ref{table_chart_plan}, four research questions are answered in this section.

\subsubsection{RQ1: What countermeasures were proposed or evaluated for in-vehicle communication systems?}
\label{S4.2.1}

\begin{figure}[ht]
\centering\includegraphics[width=1.0\linewidth]{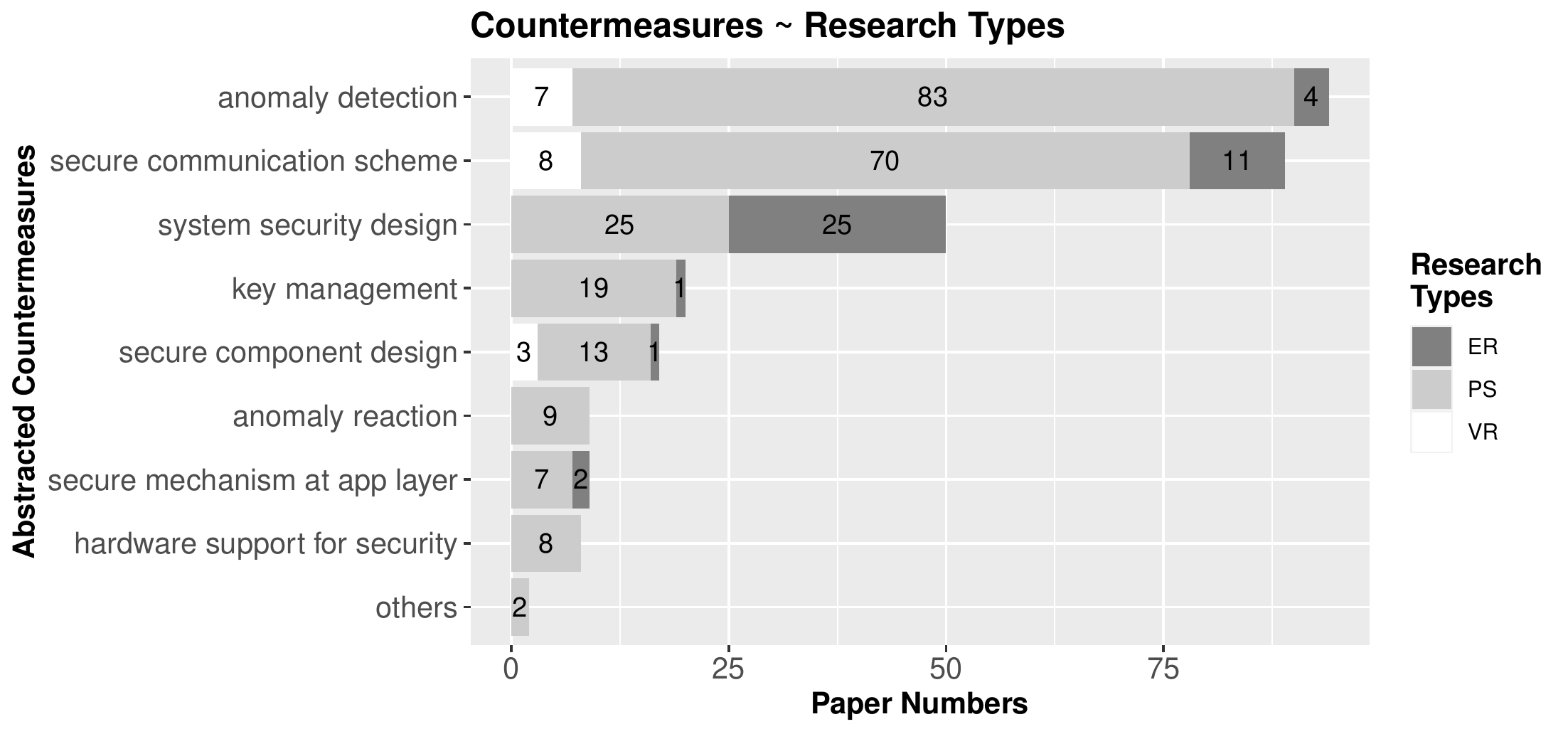}
\caption{Chart (3.1) of Countermeasures, Research Types, and Paper number}
\label{plot_3_1_coun_restype}
\end{figure}

To identify which countermeasures are mostly focused on by researchers, we abstracted the main countermeasures from publications and counted the paper numbers. The anomaly detection” is the most popular topic among studies (94), followed by the “secure communication scheme” (89). Figure \ref{plot_3_1_coun_restype} shows all numbers of abstracted countermeasures with the research type distribution. Note that there may be a bias in the number of the “secure mechanism at app layer” class because authors may use other application-specific words (e.g. vehicle diagnostic) other than general keywords used in our search string. 

For a mapping study, such an abstraction is sufficient and efficient to provide the research overview and be a preliminary study before a systematic literature review on a more specific topic (e.g. anomaly detection solutions) \cite{petersen2015guidelines}. Obscuring information can help to reduce the understanding and classification bias in a mapping.

\begin{figure}[ht]
\centering\includegraphics[width=1.0\linewidth]{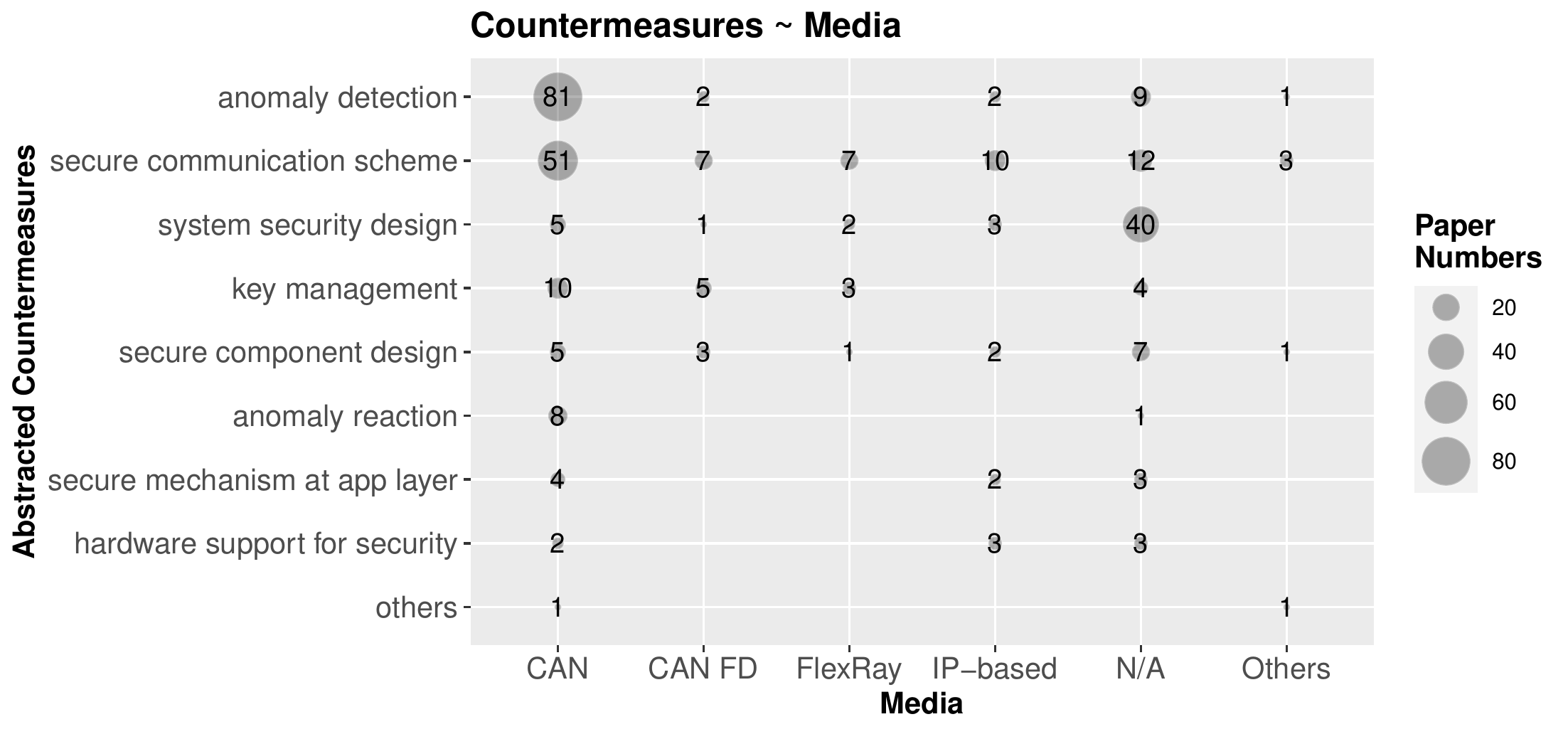}
\caption{Chart (3.2) of Countermeasures, Media types, and Papers Numbers}
\label{plot_3_2_coun_media}
\end{figure}

The publication map between countermeasures and media types is presented in Figure \ref{plot_3_2_coun_media}. CAN bus is the majority media type in the topic anomaly detection” and “secure communication scheme”. In the “system security design” topic, most research is not specific for a type of media. 

\begin{figure}[ht]
\centering\includegraphics[width=1.0\linewidth]{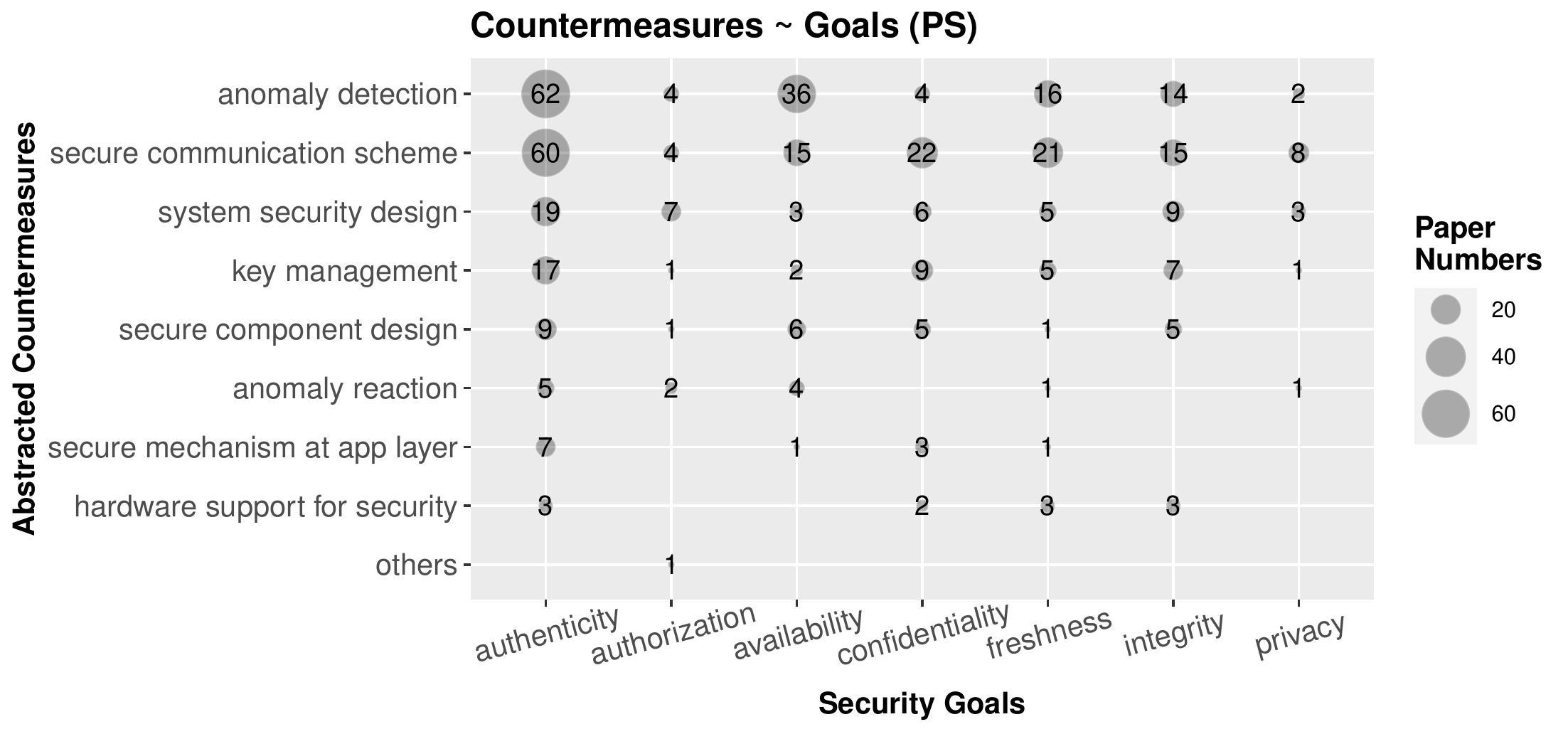}
\caption{Chart (5.1) of Countermeasures, Main Security Goals, and Numbers of Papers}
\label{plot_5_1_coun_goal}
\end{figure}

Additionally, we identified main security goals of PS papers to show which security attributes are the most concerned by researchers, which also indirectly reflects the research focuses in this field. The “authenticity” is the most referred security goal (182), followed by “availability” (67). Figure \ref{plot_5_1_coun_goal} shows the detailed mapping between countermeasures and security goals. Note that there may be a bias of the number of the “privacy” goal. In some classification scheme, privacy is not counted as a security attribute. However, we tended to use the EVITA classification scheme \cite{ruddle2009deliverable} rather than common C.I.A. triad because the EVITA one reveals more details. 

\subsubsection{RQ2: How were the proposed or existing countermeasures evaluated?}
\label{S4.2.2}

\begin{figure}[ht]
\centering\includegraphics[width=1.0\linewidth]{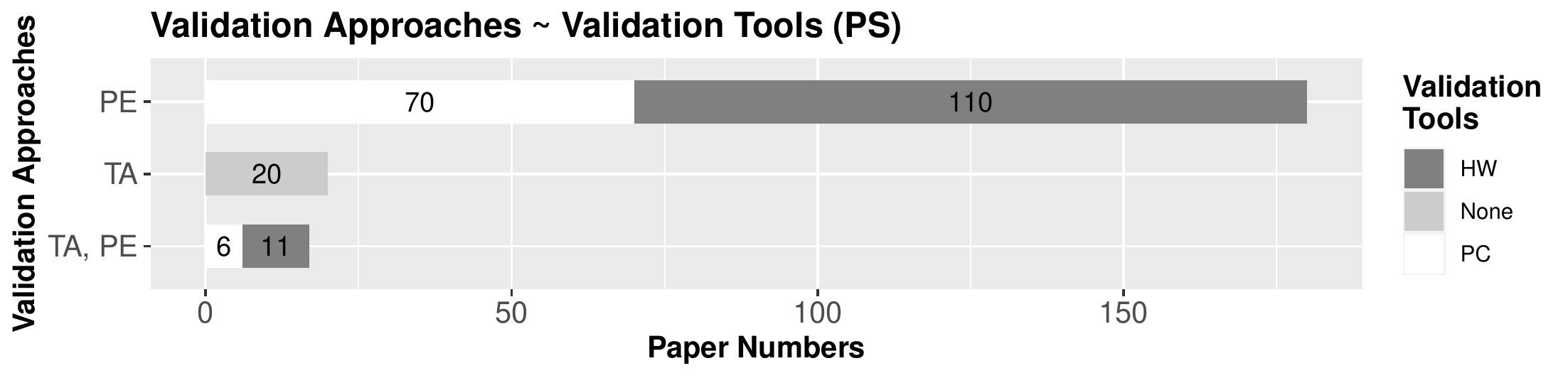}
\caption{Chart (4.1) of PS Validation Approaches, Validation Tools, and Numbers of Papers}
\label{plot_4_1_PS_valAppr_valTool}
\end{figure}

\begin{figure}[ht]
\centering\includegraphics[width=1.0\linewidth]{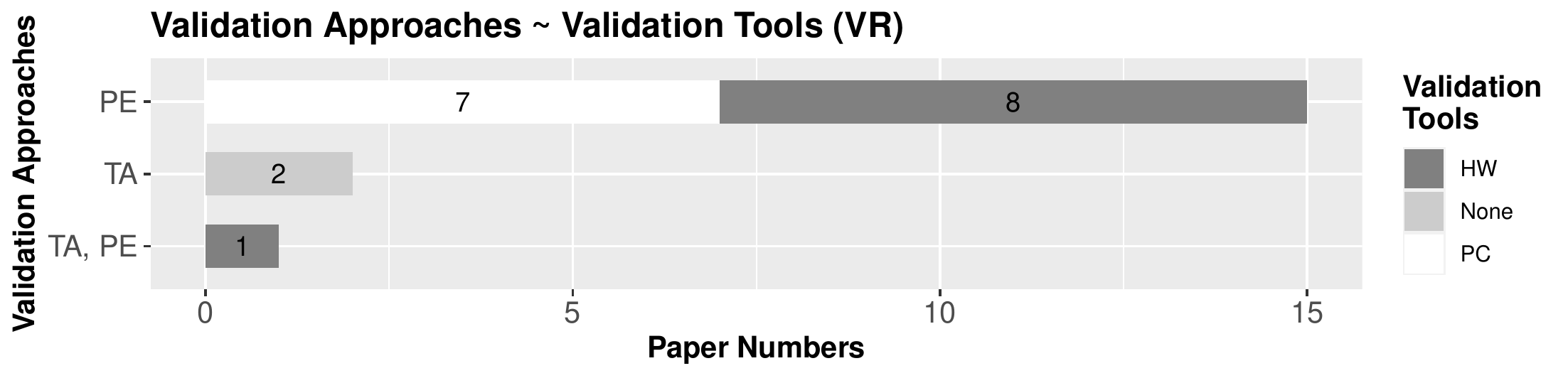}
\caption{Chart (4.2) of VR Validation Approaches, Validation Tools, and Numbers of Papers}
\label{plot_4_2_VR_valAppr_valTool}
\end{figure}

We abstracted and classified validation or evaluation methods from PS and VR papers to identify how researchers evaluate the proposed or existing solutions. Figure \ref{plot_4_1_PS_valAppr_valTool} and \ref{plot_4_2_VR_valAppr_valTool} show the distributions of PS and VR papers respectively. The majority of researchers prototyped and conducted experiments to validate their solutions. A small number of authors validated the concepts purely by theoretical analysis or mathematical proof. About 61$\%$ of valuation involved hardware devices in the validations, and the rest only used PCs to stimulate and conduct experiments. 

\begin{figure}[ht]
\centering\includegraphics[width=1.0\linewidth]{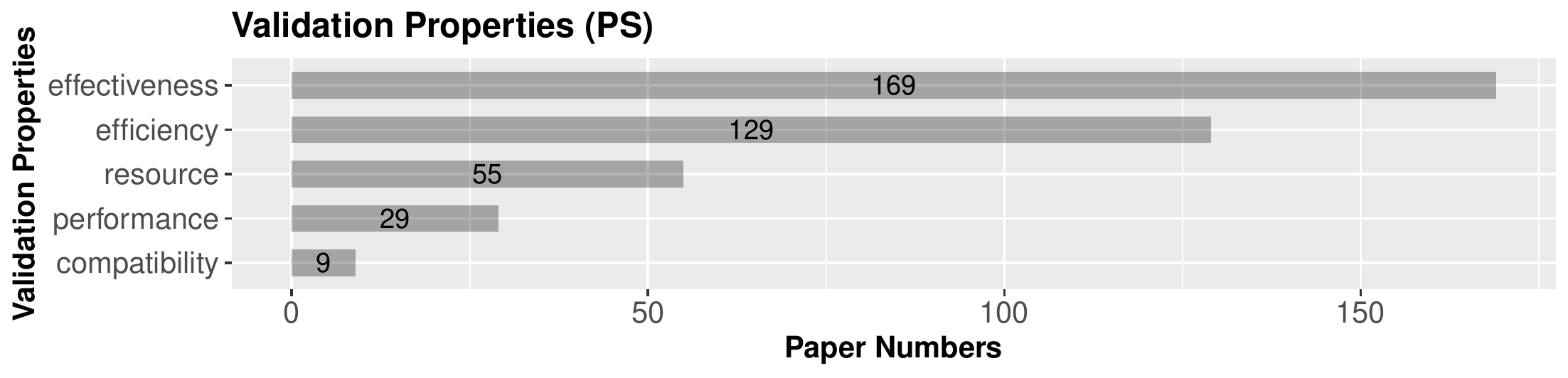}
\caption{Chart (4.3) of PS properties concerned and numbers of papers}
\label{plot_4_3_PS_valPropert}
\end{figure}

\begin{figure}[ht]
\centering\includegraphics[width=1.0\linewidth]{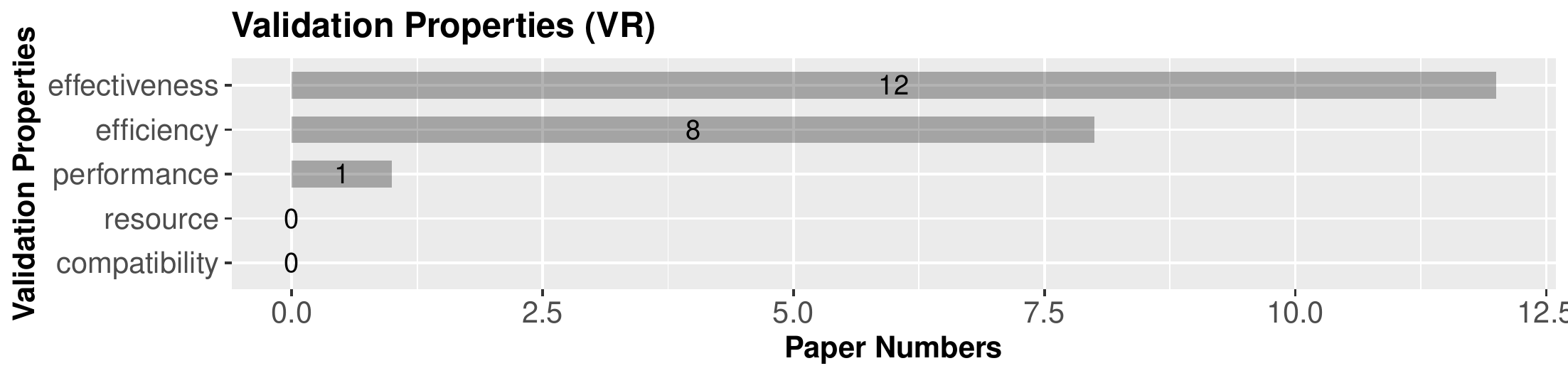}
\caption{Chart (4.4) of VR properties concerned and numbers of papers}
\label{plot_4_4_VR_valPropert}
\end{figure}

We also counted the properties concerned in the validation of PS and VR papers (shown in Figure \ref{plot_4_3_PS_valPropert} and \ref{plot_4_4_VR_valPropert}). The “effectiveness” (181) and “efficiency” (137) are the two top properties concerned in papers, followed by “resource” (55) and “performance” (30).

Additionally, we counted the total numbers of three research types (i.e. PS, VR, and ER papers) and found that 18 out of 279 papers are VR, which means that only a few researchers validate or compare the existing solutions proposed by others. Most researchers (217 PS papers) validated the proposed solutions by themselves, and about 16$\%$ (44 papers) are review papers, which are mainly based on the published literature. 

\subsubsection{RQ3: When and where were related studies conducted and published?}
\label{S4.2.3}

\begin{figure}[ht]
\centering\includegraphics[width=1.0\linewidth]{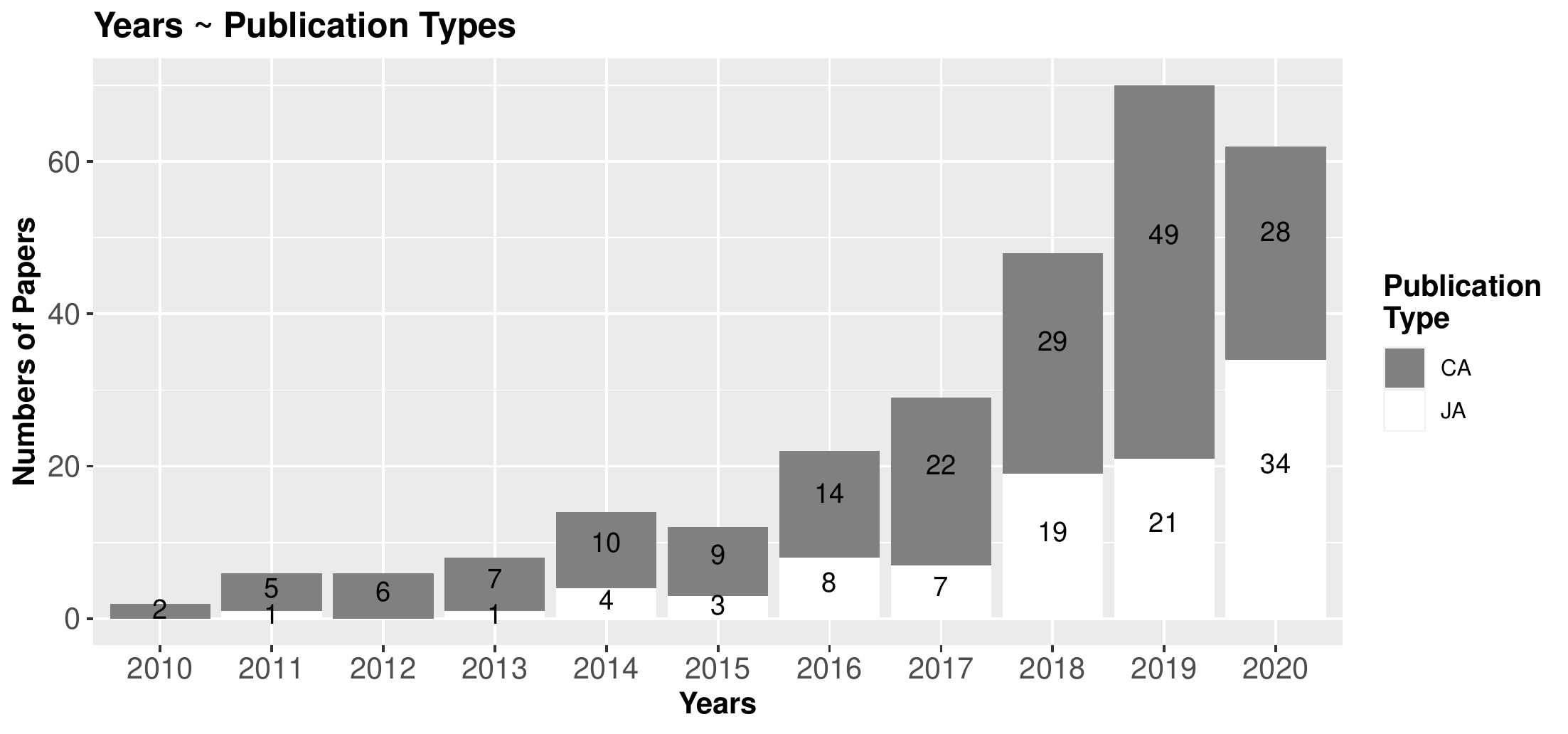}
\caption{Chart (1.1) of Numbers of Papers, Publication Types, and Years}
\label{plot_1_1_pubType_year}
\end{figure}

Figures with the keywords “year” and “country” display the paper distribution related to this research question. Figure \ref{plot_1_1_pubType_year} shows that the number of papers has been increasing rapidly in recent five years. More papers were published in conferences than in journals except the year 2020. Due to the 2020 pandemic, the overall publication number decreased and the conference papers are less than the journal ones because the pandemic had a more negative influence on conferences than journal publication. However, the journal papers did continue to increase in 2020. 

\begin{figure}[ht]
\centering\includegraphics[width=1.0\linewidth]{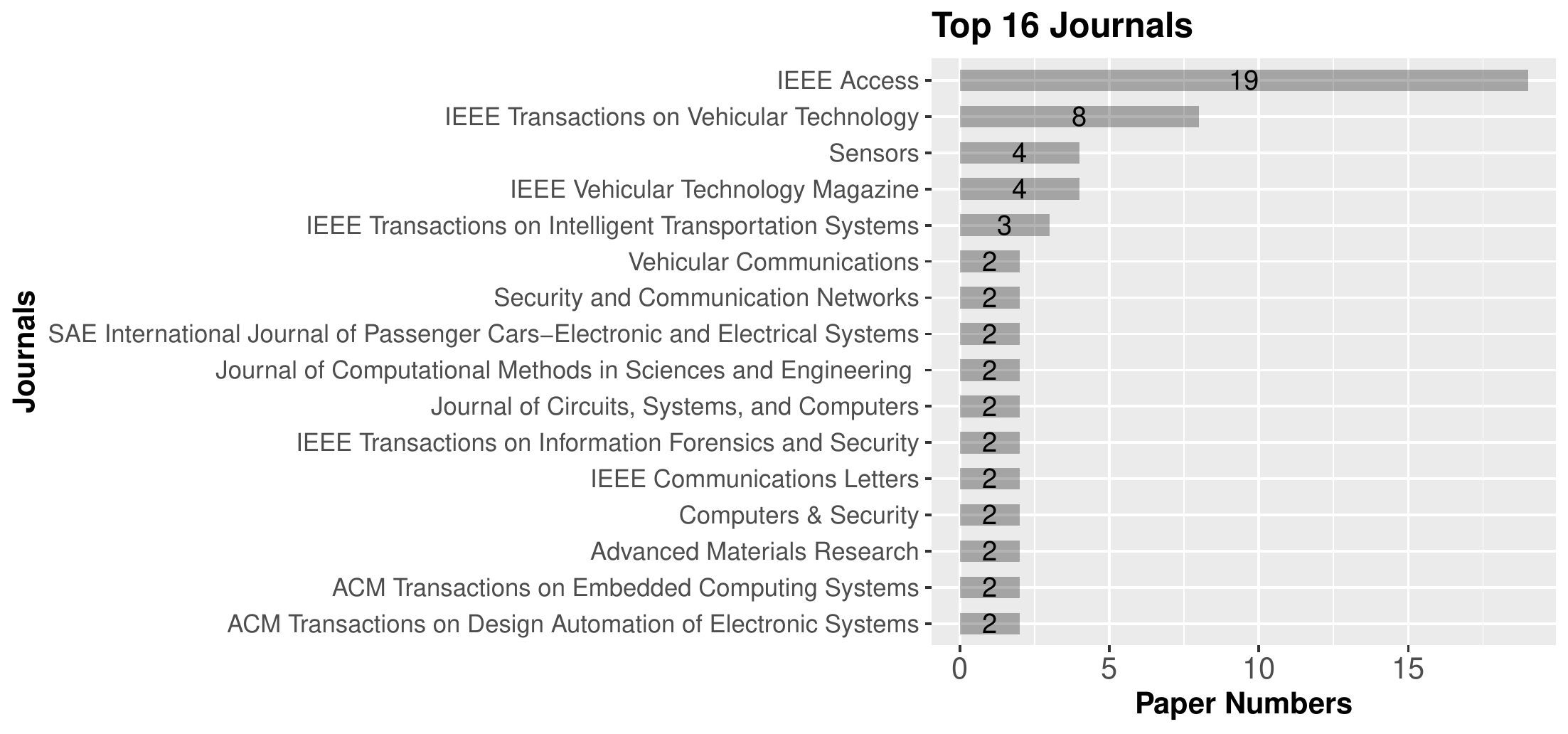}
\caption{Chart (1.2) of Top Journals and Numbers of Papers}
\label{plot_1_2_journal}

\end{figure}

\begin{figure}[ht]
\centering\includegraphics[width=1.0\linewidth]{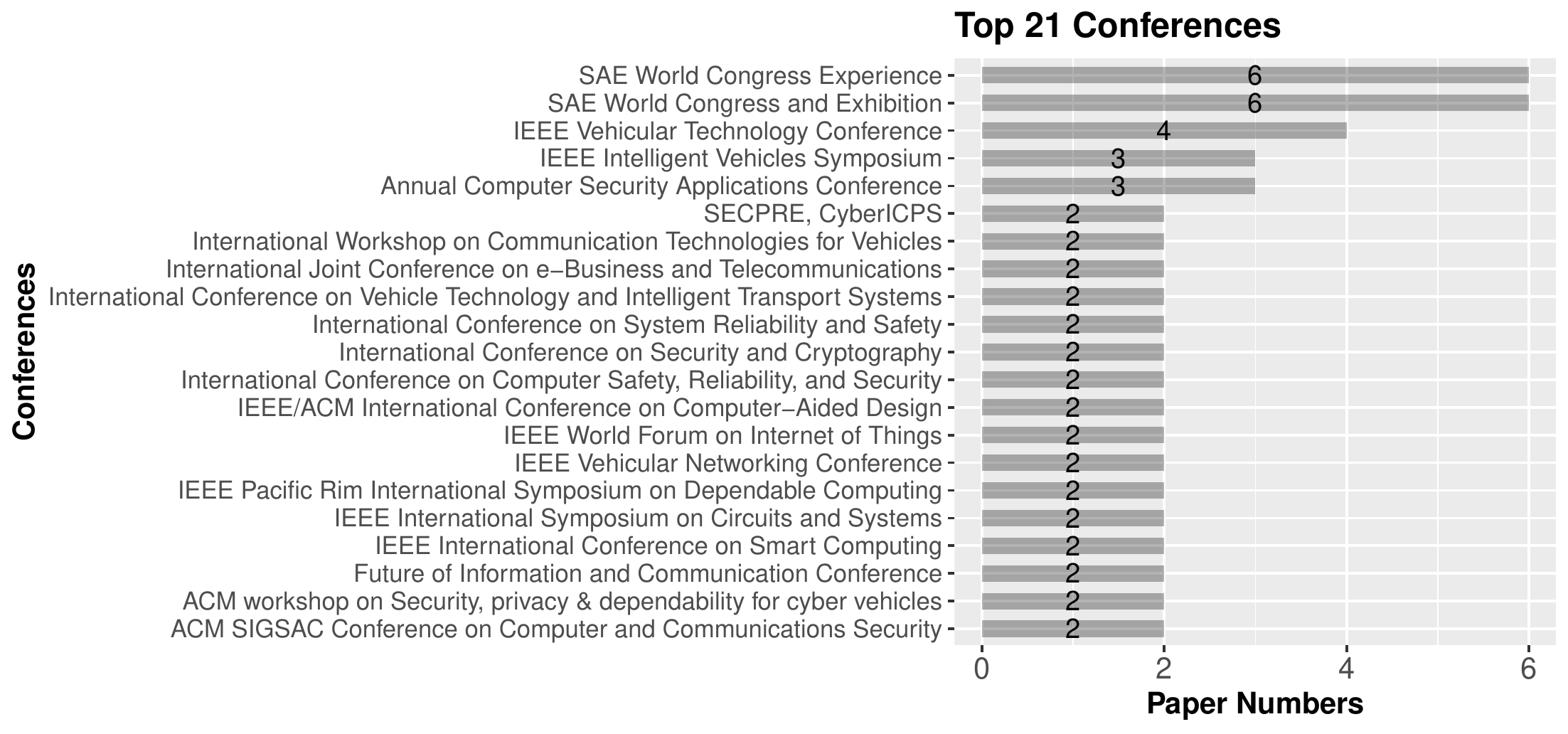}
\caption{Chart (1.3) of Top Conferences and Numbers of Papers}
\label{plot_1_3_conference}
\end{figure}

The top four journals for publications are “IEEE Access“, “IEEE Transactions on Vehicular Technology”, “IEEE Vehicular Technology Magazine”, and “Sensors”. The top three conferences for publications are “SAE World Congress Experience”, “SAE World Congress and Exhibition”, and “IEEE Vehicular Technology Conference”. More top journals and conferences can be found in Figure \ref{plot_1_2_journal} and \ref{plot_1_3_conference}. 


\begin{figure}[ht]
\centering\includegraphics[width=1.0\linewidth]{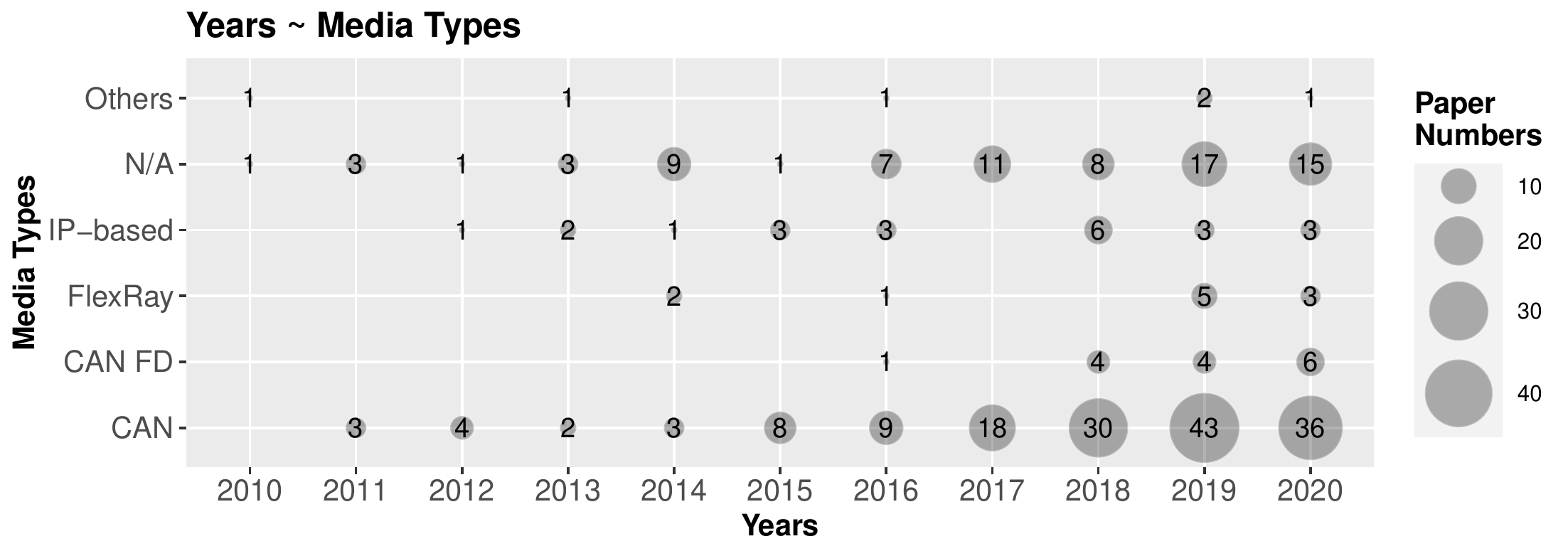}
\caption{Chart (1.4) of Media Type, Years, Numbers of Papers}
\label{plot_1_4_media_year}
\end{figure}

To view the research trend of various media types, we counted the paper numbers referring to different media (shown in Figure \ref{plot_1_4_media_year}) and found that the major media type is CAN bus, which has increased dramatically in recent years. Paper numbers of other media types (i.e. CAN FD, FlexRay, and IP-based) are not relatively high but also have increased recently compared to the early years. Paper with no certain media (i.e. “N/A”) is the second major type in the recent five years and is a non-negligible part of the target set.

\begin{figure}[ht]
\centering\includegraphics[width=1.0\linewidth]{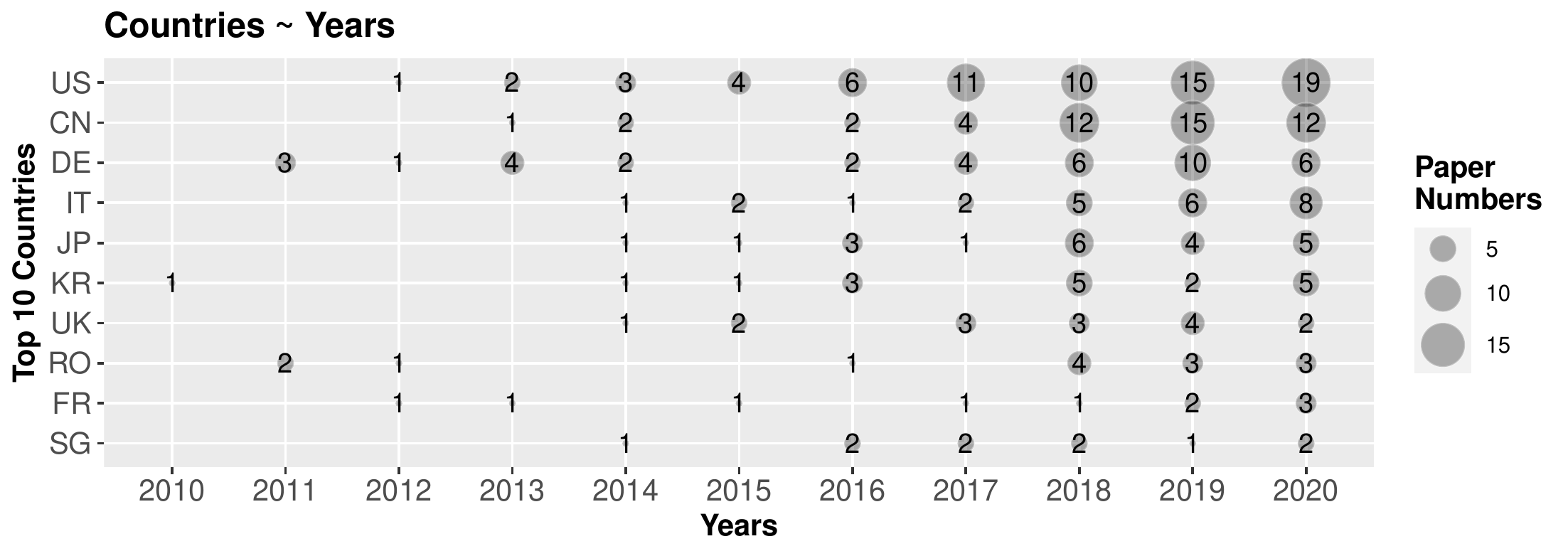}
\caption{Chart (2.1) of Top 10 Countries, Years, Numbers of Papers}
\label{plot_2_1_country_year}
\end{figure}

To answer the “where” question, we counted the publication numbers of various countries in various years. The United States (71 papers), China (48 papers), and Germany (38 papers) are the three top countries on the publication numbers. The distribution of the top ten countries is shown in Figure \ref{plot_2_1_country_year}. 

\begin{figure}[ht]
\centering\includegraphics[width=1.0\linewidth]{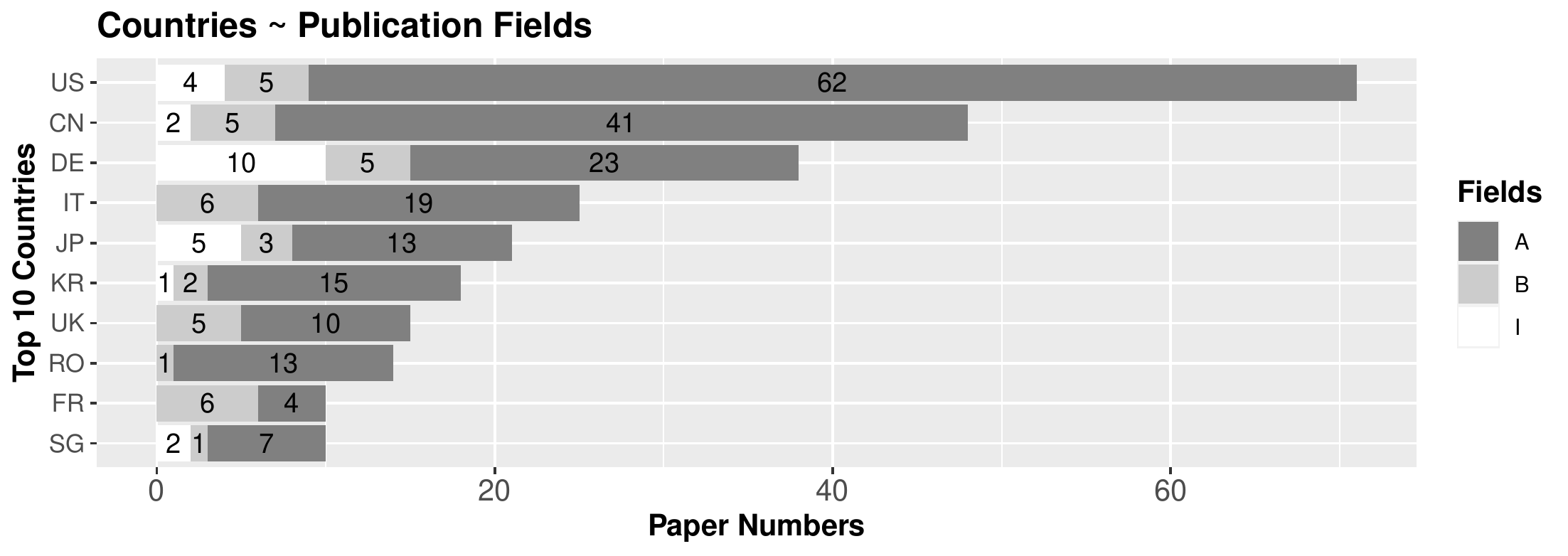}
\caption{Chart (2.2) of Numbers of Papers, Authors’ Fields, Top 10 Countries}
\label{plot_2_2_country_field}
\end{figure}

Besides, Figure \ref{plot_2_2_country_field} shows the publication fields of different countries. The majority of the research comes from the academic field. As a topic closely related to the automotive industries, each country in the figure had more or fewer corporation with industries. The United States has the top number of papers from academia, while Germany has the top number related to industrial parties.

\begin{figure}[ht]
\centering\includegraphics[width=1.0\linewidth]{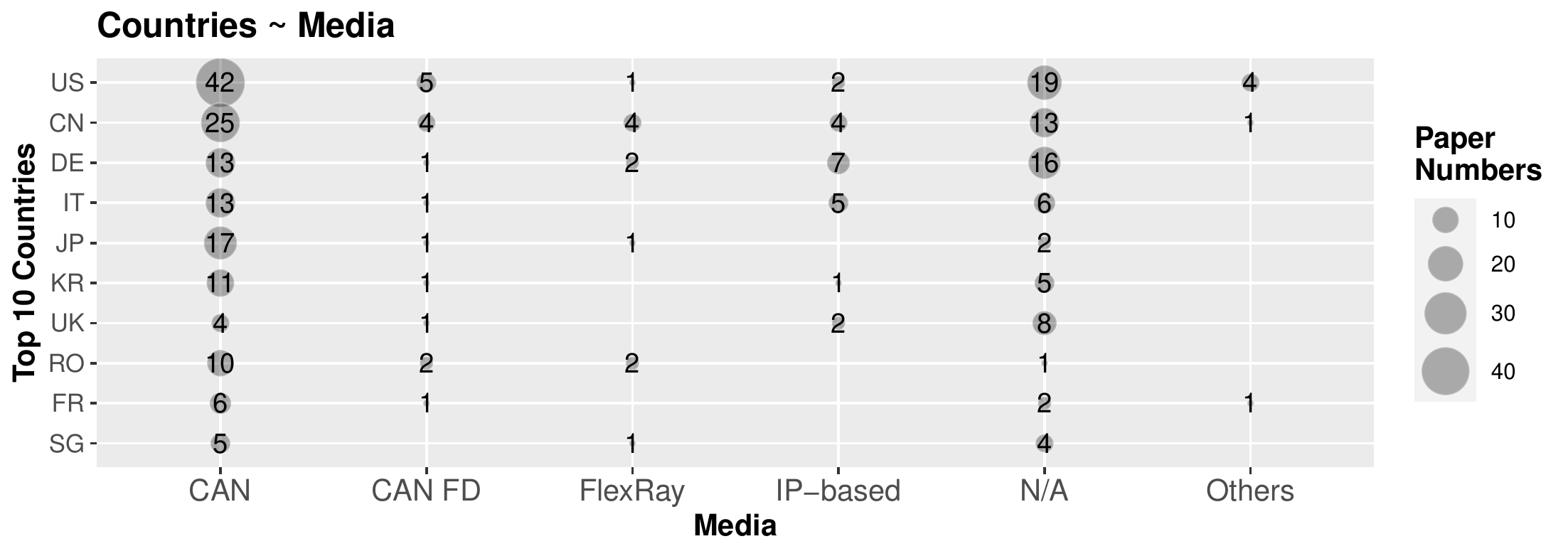}
\caption{Chart (2.3) of Numbers of Papers, Authors’ Fields, and Top 10 Countries}
\label{plot_2_3_country_media}
\end{figure}

Finally, we analyzed the media type distribution of countries, which is shown in Figure \ref{plot_2_3_country_media}. The United States has the top paper numbers on “CAN”, “CAN FD”, “N/A”, and “Others”. China and Germany have the top numbers on “FlexRay” and “IP-based” respectively. 

\subsubsection{RQ4: What are the research trends and gaps in this topic area?}
\label{S4.2.4}

\begin{figure}[ht]
\centering\includegraphics[width=1.0\linewidth]{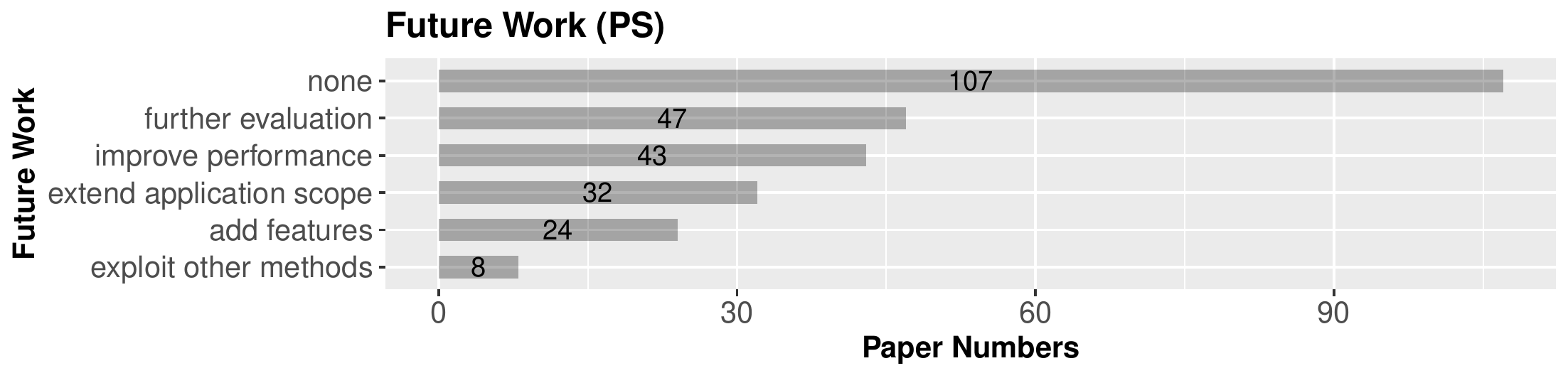}
\caption{Chart (5.2) of Future Work and Numbers of Papers}
\label{plot_5_2_PS_futurework}
\end{figure}

First, we abstracted the future work mentioned by authors in papers (shown in Figure \ref{plot_5_2_PS_futurework}). “further evaluation” (47) and “improve performance” (43) are two main kinds of future work planned by researchers, followed by “extend application scope” (32) and “add features” (24). 

Then, we synthesized all obtained results and concluded that the research interest about this topic is increasing rapidly in recent years. But the publication number drops in 2020, which may be caused by the Corona-19 pandemic in this year.

Three gaps have been identified from the data. First, although the automotive Ethernet is a hot candidate for in-vehicle backbone recently, the publication number on this media type is not high and hasn’t increased significantly until now. Most research still focused on the CAN bus, which was first proposed in 2004 and is widely used in the industry nowadays. Therefore, more research is required in the automotive Ethernet field to accelerate the implementation of this new in-vehicle network type in industry. Second, as a topic closely related to the industry, cooperation between academia and industries is necessary to promote the technology and solve real-world problems. But the cooperation number is not high so far. Third, the major research interests fall in a few certain research topics (i.e. the anomaly detection and secure communication scheme) and mainly on the CAN bus. The rest topics and media types, along with various security goals, are also important and worth being studied further.

\section{Study Evaluation}
\label{S5}
In this section, we discuss the validity threat to the study, evaluate the identified publication set, and finally present the evaluation of the overall mapping process.

\subsection{Validity threat discussion}
\label{S5.1}
The discussion of the validity threat is an essential step to analyze the quality of the mapping study via inspecting potential biases in the process. We follow the validity classification scheme used in the guideline, which includes descriptive validity, theoretical validity, generalizability, interpretive study, and repeatability \cite{petersen2015guidelines}.

\subsubsection{Descriptive validity}
\label{S5.1.1}
The descriptive validity is concerning whether the observations are described accurately and objectively \cite{petersen2015guidelines}. Although the data extraction is done by different reviewers, which may lead to descriptive biases, we developed the extraction form with detailed explanations to reduce such biases. Enumerated values of each extraction items are defined according to commonly-used schemes. Unenumerated values are abstracted and classified into certain classes through discussion. All observations are presented by a carefully-defined scheme. Therefore, the descriptive validity threat can be considered as low. 

\subsubsection{Theoretical validity}
\label{S5.1.2}
The theoretical validity is determined by the ability of being able to capture what a researcher intends to capture \cite{petersen2015guidelines}. Following biases are discussed for this study.

The sampling bias is the quality of the paper searching. We used the database search strategy and chose IEEE Xplore and ACM Digitial Library as well as two indexing databases (i.e. Web of Science and Engineer Village) in the study, which is claimed to be sufficient \cite{petersen2015guidelines}. We also added one more database called SAE MOBILUS, which is host by the Society of Automotive Engineers (SAE) and specific for the automotive domain. Search strings were discussed by reviewers and considered as enough to hit the target topic. However, if a paper is related to the topic but does not use any word in the search string in its title or abstract, this paper may be missed. We didn’t perform the forward and backward snowballing due to the trade-off between the sampling numbers and workload since we’ve already got almost 300 records after the primary inclusion and exclusion process. It is normally not possible to includes all papers in a study. Wohlin et al. \cite{wohlin2013reliability} also pointed out that even two mapping studies of the same topic ended up with different sets of articles. Since our purpose of this mapping study is to provide an overview of the target topic instead of identifying all related papers and collecting all evidence from the population, it is considered acceptable to do the sampling via only the database search. 

The researcher bias is related to the researchers’ ability and the quality of their works. To reduce such validity threats, proper work processes were defined based on the widely-used guideline \cite{petersen2015guidelines}, which has been cited more than 1000 times by the point we wrote this report. For the inclusion and exclusion process, two reviewers made the decision separately and solved conflicts by meetings. A third reviewer made the final decision for the confusing records. We also kept a conservative style in the work. If a reviewer is not 100 percent sure whether a paper should be included or excluded, it would be included in this round and would be discussed or checked with the full text later to ensure that no related paper is excluded wrongly. For the data extraction, an extraction form with detailed value definitions is developed as explained previously. Meetings were held to discuss and introduce the extraction strategies. A pilot data extraction was also conducted first with 10 random papers to unify the understanding and extraction style of different reviewers. A final check was performed by the 1st reviewer, who leaded the data extraction process.

The publication bias represents the bias between the publication state and the real research state in a certain topic area. First, negative or controversial results may not be published. Novel and valuable techniques in industries may also not be published due to commercial considerations. Second, papers not written in English are excluded but they may also include related ideas. Third, authors' publication behaviors affect the number of papers. Authors that have enough research time in a project may publish several papers on the same idea with different emphasis, while others that have less time can publish only one paper on one approach. Sometimes a journal paper is also published as an extension of a conference paper. Fourth, since we excluded papers whose full text is not available for us, some relevant papers are missing. Fifth, although we aim to present the current state of the target research area, there is an inevitable time lag from finishing the research to the real publication time.

Although many efforts have been done to reduce the mentioned theoretical validity threats, it is still not possible to be eliminated due to the human judgment involved in the study and the inevitable biases of publications.

\subsubsection{Generalizability}
\label{S5.1.3}
The result generalizability of the mapping involves both internal and external generalizability. The internal one is the generalization within the population. Since we conducted the study following the widely-used guideline with detailed working protocols and recordings, the internal generalization threat is low. The external one is the generalization between different populations. The results of this study can not be applied to other similar topics (e.g. security countermeasures in other systems or industries) since the technical legacy and context are completely different.

\subsubsection{Interpretive validity}
\label{S5.1.4}
The interpretive validity, or called conclusion validity, is about whether the conclusions are reasonable given the data. A threat of such validity is due to researcher bias (already discussed in the theoretical validity). The background, knowledge, and experience of reviewers will cause such interpretive bias. Detailed working protocols are defined to specify the study behaviors and minimize the threats caused by personal subjective judgment. Besides, previous experience with mapping studies may also increase the interpretive validity. Although it is the first time for the 1st, 3rd, and 4th authors to conduct an SMS, the 2nd author is experienced with mapping studies and systematic literature reviews, who guided and reviewed the study finally.

\subsubsection{Repeatability}
\label{S5.1.5}

The repeatability is guaranteed by following a well-defined work protocol and reporting the study process in detail. This article elaborates on all necessary work specifications and results to make this study repeatable.

\subsection{Study identification evaluation}
\label{S5.2}
To evaluate the quality of the study identification results, we compared the final included papers to a known set, which has been prepared before the study and includes relevant papers we already knew in previous research. Table \ref{table_known_papers} shows the list of known papers and the comparison results. 8 out of 14 known papers are covered by the final included paper set but 6 are not hit by the identification process. We analyzed the result and tried to find reasons for missing papers. We found that all missing papers mainly focus on particular techniques (No. 4 is on the firewall; No. 10 is on the secure gateway, No. 6 to 8 are on intrusion detection) or concrete application scenarios (No. 11 is about diagnostics in repair shops) and do not use the common words we picked for the search string. This searching bias has been mentioned in the theoretical validity discussion. However, on the other hand, if we do not use more general words like “security countermeasure” in the initial search, we may have no idea what are the specific countermeasures that could be used in this field.

To check how many papers we can newly find if we use more specific words in the search, we designed a supplementary search string, which is “((anomaly detection) OR (intrusion detection) OR (secur* gateway) OR (secur* diagnostic) OR firewall) AND ((in-vehicle OR automotive OR automobile) AND (network OR communication))”. By using more specific words, we found an additional huge number of records (2096). To balance the workload and the size of the collection, we do not add the records found by specific words into the initial collection. Therefore, the papers we identified are those that discuss the security issues using more general words. If the researchers want to achieve an overview of a specific technique (e.g. anomaly detection approaches), they can narrow the scope down to conduct an SMS or SLR on the particular technique with an acceptable workload and a relatively complete collection.

Applying snowballing is also a good choice to migrate bias caused by the search words since it can found content-related papers regardless of the keywords used. However, snowballing requires a huge amount of workload especially starting from a big paper set. Reviewers should make a trade-off decision between the size of sampling and the work efficiency. Since the topic in this SMS is a general one and our purpose of this study is to provide an overview of this topic area instead of finding all evidence, we do not use snowballing in this mapping. In further possible SLRs, the forward and backward snowballing is a good supplementary search strategy to find all evidence on a more specific topic with an acceptable workload. 

Note that we don’t add the missing known papers into the included paper set because they were found by manual search randomly and not consistent with the paper identification process in this study.

\begin{table}[ht]
\renewcommand{\arraystretch}{1.1}
\caption{Known Paper Set and Comparison Results}
\label{table_known_papers}
\centering
\begin{tabular}{p{0.5cm} p{9.5cm}p{1cm}}
\hline
\textbf{No.} & \textbf{Paper Title} & \textbf{If Hit}\\
\hline
1 & CAN Security: Cost-Effective Intrusion Detection for Real-Time Control Systems Overview of In-Vehicle Networks \cite{otsuka2014can} & Yes \\
2 & Comparative performance evaluation of intrusion detection methods for In-Vehicle networks \cite{ji2018comparative} & Yes \\
3 & Experimental Evaluation of Cryptography Overhead in Automotive Safety-Critical Communication \cite{junior2018experimental} & Yes \\
4 & Hardware/Software Co-Design of an Automotive Embedded Firewall \cite{pese2017hardware} & No \\
5 & Integrated Safety and Security Development in the Automotive Domain \cite{macher2017integrated} & Yes \\
6 & Intrusion detection in connected cars \cite{haas2017intrusion} & No \\
7 & Intrusion prevention system of automotive network CAN bus \cite{abbott2016intrusion} & No \\
8 & Research on CAN Network Security Aspects and Intrusion Detection Design \cite{li2017research} & No \\
9 & Review of Secure Communication Approaches for In-Vehicle Network \cite{hu2018review} & Yes \\
10 & Secure automotive gateway - Secure communication for future cars \cite{seifert2014secure} & No \\
11 & Securing vehicle diagnostics in repair shops \cite{kleberger2014securing} & No \\
12 & Security Mechanisms Design for In-Vehicle Network Gateway \cite{luo2018security} & Yes \\
13 & Security mechanisms design of automotive gateway firewall \cite{luo2019security}& Yes \\
14 & VoltageIDS: Low-level communication characteristics for automotive intrusion detection system \cite{choi2018voltageids} & Yes \\
\hline
\end{tabular}
\end{table}

\subsection{Mapping process evaluation}
\label{S5.3}
Finally, we evaluate the mapping process by an evaluation rubric devised in the guideline \cite{petersen2015guidelines}. A total of 26 relevant activities that can be applied in mapping studies have been identified. The guideline suggests calculating the ratio of the number of actions applied comparing to the total number. Table \ref{table_activity_conducted} shows the total activities and the applied ones in this study. 14 activities were conducted and the ratio is 54$\%$. According to the study \cite{petersen2015guidelines} on the SMS conducted before 2015, the maximum and median ratio is about 48$\%$ and 33$\%$. Therefore, the quality of this study can be considered as high. 

\begin{table}[ht]
\renewcommand{\arraystretch}{1.1}
\caption{Activities Conducted in this Study}
\label{table_activity_conducted}
\centering
\begin{tabular}{p{1.8cm} p{6.5cm} p{2.8cm} }
\hline
\textbf{Phase} & \textbf{Activity} & \textbf{If Applied (Where)} \\
\hline
\multirow{3}{1.5cm}{Need for map} & Motivate the need and relevance & Yes (Sec. \ref{S1})\\
\cline{2-3} & Defined objectives and questions & Yes (Sec. \ref{S3.1}) \\
\cline{2-3} & Consult with target audience to define questions& No  \\
\hline
\multirow{3}{2cm}{Choosing search strategy} & Snowballing & No \\
\cline{2-3} & Manual & No \\
\cline{2-3} & Conduct database search & Yes (Sec. \ref{S3.2}, \ref{S3.3})  \\
\hline
\multirow{4}{1.5cm}{Develop the search} & PICO & No \\
\cline{2-3} & Consult librarians or experts & No \\
\cline{2-3} & Iteratively try finding more relevant papers & Yes (Sec. \ref{S3.2.1}) \\
\cline{2-3} & Use standards, encyclopedias, and thesaurus & No \\
\hline
\multirow{4}{1.5cm}{Evaluate the search} & Test-set of known papers & Yes (Sec. \ref{S5.2}) \\
\cline{2-3} & Expert evaluates results & No \\
\cline{2-3} & Search web-pages of key authors & No \\
\cline{2-3} & Test-retest & No \\
\hline
\multirow{3}{1.5cm}{Inclusion and Exclusion} & Identify objective criteria for decision & Yes (Sec. \ref{S3.2.2}) \\
\cline{2-3} & Add additional reviewer, resolve disagreements between them when needed & Yes (Sec. \ref{S3.3}) \\
\cline{2-3} & Decision rules & Yes (Sec. \ref{S3.3}) \\
\hline
\multirow{4}{1.5cm}{Extraction process} & Identify objective criteria for decision & Yes (Section \ref{S3.4.2}) \\
\cline{2-3} & Obscuring information that could bias & Yes (Sec. \ref{S3.5}) \\
\cline{2-3} & Add additional reviewer, resolve disagreements between them when needed & Yes (Sec. \ref{S3.4.1}) \\
\cline{2-3} & Test-retest & No \\
\hline
\multirow{3}{1.5cm}{Classification scheme} & research type & Yes (Sec. \ref{table_TopicIndExtraTemplate}) \\
\cline{2-3} & research method & No \\
\cline{2-3} & Venue type & Yes (Sec. \ref{table_TopicIndExtraTemplate}) \\
\hline
Validity discussion & Validity discussion, limitations provided & Yes (Sec. \ref{S5.1}) \\
\hline
\end{tabular}
\end{table}

\section{Conclusion}
\label{S6}

In this mapping study, we first worked out the systematic workflow according to a widely-used guideline and defined strategies and criteria of study identification and data extraction steps. By following the work protocols, we identified 279 related papers, extracted data from the metadata and full text of selected papers, and visualized the abstracted and classified data by (stacked) bar or bubble charts. Finally, we discussed the validity threats of this mapping and evaluated the identified paper set as well as the whole study process.

We mainly focus on four research questions in this study. For the RQ1, security countermeasures are collected and classified into nine classes, in which the anomaly detection” and “secure communication scheme” are the two top kinds of countermeasures discussed by researchers. Most of the countermeasures are mentioned in the PS papers, and the majority of them are specific to the CAN bus networks. Authenticity is the most mentioned security goal by researchers. For the RQ2, the majority of the validation approach is “prototyping and then doing experiments”, in which about 61\% of validation processes require hardware devices other than computers. The effectiveness and efficiency are the two top properties concerned in the validation. For the RQ3, the publication numbers have been increased rapidly in recent years, but there’s a drop in 2020 due to the possible impact of the Corona-19 pandemic. The publication distribution of media types in different years is also displayed in a bubble chart to show the trends of each type. The top three countries in this field are the United States, China, and Germany, and the majority of research comes from the academic field. The distribution of various media types related to countries is presented to show which country contributes the most to a certain media type. For the RQ4, the future work mentioned in papers is counted and visualized, which shows that “further evaluation” is the most mentioned future work by authors. 

Finally, we concluded the trends and gaps based on all data we obtained in this study. The research interest on security issues has been increasing rapidly in recent years, but the publication number dropped in 2020 due to the pandemic (the most possible reason). Three research gaps are identified. First, there’s not much research related to the Ethernet, which has attracted much attention in the automotive industry recently. Second, the cooperation between academia and industry should be enhanced to solve real-world problems and accelerate the technique evolution. Third, the majority of research in this area focuses on a few subtopics (e.g. anomaly detection, protecting the CAN bus), other countermeasures and media types with less attention (e.g. anomaly reaction, protecting the automotive Ethernet networks) are still worth being studied.

The outcomes of this study, including an inventory of relevant papers and the extracted data, are useful references for research on a certain security technique or further SLRs in this topic area. The reported workflow, strategies, and criteria can also be served as examples for SMSs on other topics. This study, not like other review articles in this research field, investigates not only the research findings (i.e. security countermeasures and relevant validation methods) but also research activity patterns (e.g. publication years, countries, and author’s fields). From such a broader perspective, researchers can have an overview of the current research state at a higher level and identify research trends and gaps for better future work.






\bibliographystyle{elsarticle-num-names}
\bibliography{ref.bib}







\end{document}